\documentclass[prd,showpacs,showkeys,floatfix,twocolumn,amsmath,amssymb,floatfix]{revtex4}
\usepackage{graphicx,color,dcolumn,booktabs,bm}
\usepackage{longtable,lscape}
\usepackage{amssymb}
\usepackage{indentfirst}
\usepackage{epsfig}
\usepackage{feynmf}   
\usepackage{epstopdf}   
\usepackage{slashed}  
\usepackage{cases}
\usepackage{color}
\usepackage{multirow}
\usepackage{graphicx,color,dcolumn,booktabs,bm}
\usepackage[colorlinks, citecolor=blue,anchorcolor=red,menucolor=red, linkcolor=red,filecolor=red,runcolor=red,urlcolor=blue,frenchlinks=red]{hyperref}

\begin{document}

\title{$D$-wave heavy baryons of the  $SU(3)$ flavor $\mathbf{6}_F$}

\author{Qiang Mao$^{1,2}$}
\author{Hua-Xing Chen$^{2}$}
\email{hxchen@buaa.edu.cn}
\author{Atsushi Hosaka$^{3,4}$}
\email{hosaka@rcnp.osaka-u.ac.jp}
\author{Xiang Liu$^{5,6}$}
\email{xiangliu@lzu.edu.cn}
\author{Shi-Lin Zhu$^{7,8,9}$}
\email{zhusl@pku.edu.cn}
\affiliation{
$^1$Department of Electrical and Electronic Engineering, Suzhou University, Suzhou 234000, China\\
$^2$School of Physics and Beijing Key Laboratory of Advanced Nuclear Materials and Physics, Beihang University, Beijing 100191, China \\
$^3$Research Center for Nuclear Physics, Osaka University, Ibaraki 567--0047, Japan \\
$^4$Advanced Science Research Center, Japan Atomic Energy Agency, Tokai, Ibaraki, 319-1195 Japan \\
$^5$School of Physical Science and Technology, Lanzhou University, Lanzhou 730000, China\\
$^6$Research Center for Hadron and CSR Physics, Lanzhou University and Institute of Modern Physics of CAS, Lanzhou 730000, China\\
$^7$School of Physics and State Key Laboratory of Nuclear Physics and Technology, Peking University, Beijing 100871, China \\
$^8$Collaborative Innovation Center of Quantum Matter, Beijing 100871, China \\
$^9$Center of High Energy Physics, Peking University, Beijing 100871, China}

\begin{abstract}
We use the method of QCD sum rules to study the $D$-wave charmed and bottom baryons of the $SU(3)$ flavor $\mathbf{6}_F$, and calculate their masses up to the order $\mathcal{O}(1/m_Q)$ with $m_Q$ the heavy quark mass.
Our results suggest that the $\Xi_c(3123)$ can be well interpreted as a $D$-wave $\Xi_c^\prime(\mathbf{6}_F)$ state, and it probably has a partner state close to it. Our results also suggest that there may exist as many as four $D$-wave $\Omega_c$ states in the energy region $3.3\sim3.5$ GeV, and we propose to search for them in the further LHCb and BelleII experiments.
\end{abstract}

\maketitle

\section{Introduction}
\label{sec:intro}

In the past years important experimental progresses have been made in the field of heavy baryons~\cite{pdg}, and we refer to reviews~\cite{Chen:2016spr,Cheng:2015iom,Crede:2013sze,Klempt:2009pi,Bianco:2003vb,Korner:1994nh} for their recent progress.
Especially, the LHCb experiment observed as many as five excited $\Omega_c$ states at the same time~\cite{Aaij:2017nav}, which is probably related to the fine structure of the strong interaction.
Some of these excited $\Omega_c$ states can be interpreted as $P$-wave states~\cite{Chen:2015kpa}, inspiring us to further study the $D$-wave heavy baryons.
Actually, in Ref.~\cite{Chen:2016phw} we have systematically studied the $D$-wave heavy baryons of the $SU(3)$ flavor $\mathbf{\bar 3}_F$ using the method of QCD sum rules~\cite{Shifman:1978bx,Reinders:1984sr} in the framework of heavy quark effective theory (HQET)~\cite{Grinstein:1990mj,Eichten:1989zv,Falk:1990yz}, and in the present study we shall follow the same approach to study the $SU(3)$ flavor $\mathbf{6}_F$ ones, including the $D$-wave $\Omega_c$ states.

There have been lots of heavy baryons observed in various experiments~\cite{pdg,Yelton:2016fqw,Kato:2016hca,Aaij:2017vbw}. Among them, the $\Xi_c(3123)$ observed by the BaBar Collaboration~\cite{Aubert:2007dt} (but not seen in the following Belle experiment~\cite{Kato:2013ynr}) is a good candidate of the $D$-wave $\Xi_c^\prime(\mathbf{6}_F)$ state, and has been investigated using many phenomenological methods/models,
including various quark models~\cite{Ebert:2007nw,Zhong:2007gp,Ebert:2011kk,Liu:2012sj,Shah:2016mig,Zhao:2016qmh},
the Faddeev method~\cite{Valcarce:2008dr},
the Regge trajectory~\cite{Guo:2008he},
the relativistic flux tube model~\cite{Chen:2014nyo},
the heavy hadron chiral perturbation theory~\cite{Cheng:2015naa},
and QCD sum rules~\cite{Zhang:2008pm,Wang:2017vtv}, etc.
More discussions can be found in Refs.~\cite{Cheng:2006dk,Selem:2006nd,Garcilazo:2007eh,Gerasyuta:2007un,Guo:2015daa}, and we again refer to reviews~\cite{Chen:2016spr,Cheng:2015iom,Crede:2013sze,Klempt:2009pi,Bianco:2003vb,Korner:1994nh} for the recent progress.

We have studied the heavy baryons using the method of QCD sum rules within HQET~\cite{Liu:2007fg,Chen:2015kpa,Mao:2015gya,Chen:2016phw},
and in the present study we shall further study the $D$-wave charmed baryons of the $SU(3)$ flavor $\mathbf{6}_F$.
We shall take the ${\mathcal O}(1/m_Q)$ corrections ($m_Q$ is the heavy quark mass) into account during our QCD sum rule analyses, and extract the chromomagnetic splitting within the same baryon multiplet.
More discussions on heavy mesons and baryons containing a single heavy quark can be found in Refs.~\cite{Bagan:1991sg,Neubert:1991sp,Broadhurst:1991fc,Ball:1993xv,Huang:1994zj,Dai:1996yw,Colangelo:1998ga,Groote:1996em,Zhu:2000py,Lee:2000tb,Huang:2000tn,Wang:2003zp,Duraes:2007te,Zhou:2014ytp,Zhou:2015ywa}.

This paper is organized as follows. In Sec.~\ref{sec:sumrule}, we briefly introduce how we use the $D$-wave heavy baryon interpolating fields to perform QCD sum rule analyses,
and we refer interested readers to consult the discussion in Refs.~\cite{Chen:2016phw,Liu:2007fg,Chen:2015kpa,Mao:2015gya} for details.
Then we perform numerical analyses in Sec.~\ref{sec:numerical}, and offer a short summary in Sec.~\ref{sec:summary}.

\section{QCD Sum Rule Analyses}
\label{sec:sumrule}

\begin{figure*}[hbt]
\begin{center}
\scalebox{0.6}{\includegraphics{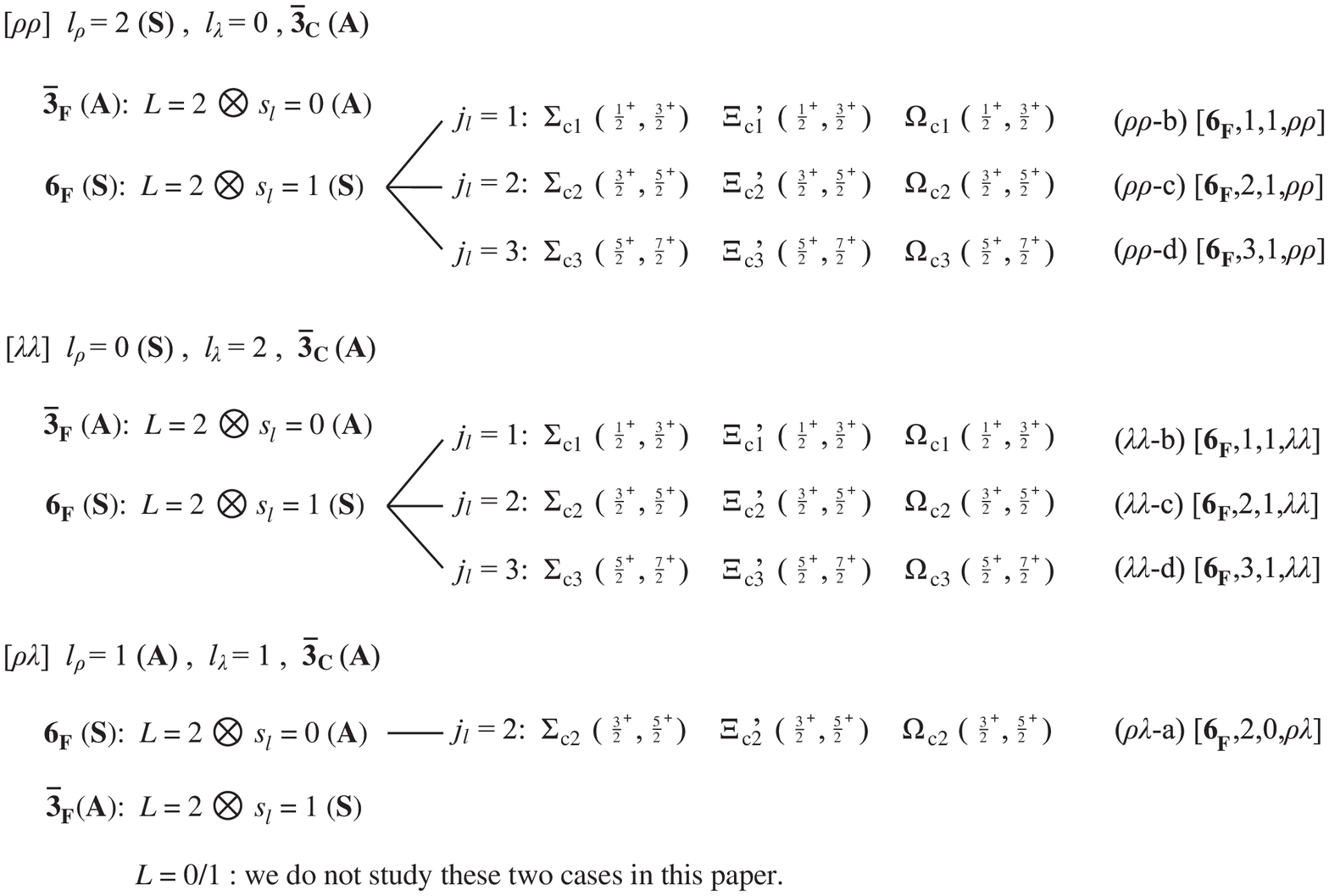}}
\end{center}
\caption{The notations for the $D$-wave charmed baryons of the $SU(3)$ flavor $\mathbf{6}_F$. See Fig.~1 of Ref.~\cite{Chen:2016phw} and discussions therein for details.
\label{fig:dwave}}
\end{figure*}

In Ref.~\cite{Chen:2016phw} we have systematically constructed all the $D$-wave heavy baryon interpolating fields, but just studied the $SU(3)$ flavor $\mathbf{\bar 3}_F$ ones using the method of QCD sum rules in the
framework of heavy quark effective theory (HQET). In this paper we use the same method to study the rest ones of the $SU(3)$ flavor $\mathbf{6}_F$, as briefly shown in Fig.~\ref{fig:dwave}. We refer interested readers to consult the discussion in Ref.~\cite{Chen:2016phw} for details.

Here we briefly explain our notations: $J^{\alpha_1\cdots\alpha_{j-1/2}}_{j,P,F,j_l,s_l,\rho-\lambda}$ denotes the $D$-wave heavy baryon field belonging to the baryon multiplet $[F, j_l, s_l, \rho-\lambda]$, where $j$, $P$, and $F$ denote its total angular momentum, parity and $SU(3)$ flavor representation ($\mathbf{\bar 3}_F$ or $\mathbf{6}_F$); $j_l$ and $s_l$ are the total angular momentum and spin angular momentum of its light components, respectively; there are three types ($\rho-\lambda$): $\rho\rho$--type ($l_\rho = 2$ and $l_\lambda = 0$), $\lambda\lambda$--type ($l_\rho = 0$ and $l_\lambda = 2$) and $\rho\lambda$--type ($l_\rho = 1$ and $l_\lambda = 1$), where we use $l_\rho$ to denote the orbital angular momentum between the two light quarks, and $l_\lambda$ to denote the orbital angular momentum between the heavy quark and the two-light-quark system. These parameters satisfy $L = l_\lambda \otimes l_\rho = 2$ (note that we only investigate the $D$-wave heavy baryons in the present study), $j_l = L \otimes s_l$ and $j = j_l \otimes s_Q = | j_l \pm 1/2 |$, with $s_Q = 1/2$ the spin of the heavy quark.

The explicit forms of $J^{\alpha_1\cdots\alpha_{j-1/2}}_{j,P,F,j_l,s_l,\rho-\lambda}$ have been given in Eqs.~(2--19) of Ref.~\cite{Chen:2016phw}. We use them to perform QCD sum rule analyses by assuming their coupling to the state $|j,P,F,j_l,s_l,\rho-\lambda\rangle$ to be
\begin{eqnarray}
&& \langle 0| J^{\alpha_1\cdots\alpha_{j-1/2}}_{j,P,F,j_l,s_l,\rho-\lambda} |j,P,F,j_l,s_l,\rho-\lambda \rangle
\\ \nonumber && ~~~~~~~~~~~~~~~~~~~~~~~~~~~~~~~~~~ = f_{F,j_l,s_l,\rho-\lambda} u^{\alpha_1\cdots\alpha_{j-1/2}} \, .
\end{eqnarray}
Again, we recommend interested readers to consult Refs.~\cite{Chen:2016phw,Liu:2007fg,Chen:2015kpa,Chen:2008qv,Mao:2015gya,Chen:2015moa,Chen:2016otp} for details, but simply list here the equation to evaluate the mass of the heavy baryon belonging to the multiplet $[F, j_l, s_l, \rho-\lambda]$:
\begin{equation}
\label{eq:mass}
m_{j,P,F,j_l,s_l,\rho-\lambda} = m_Q + \overline{\Lambda}_{F,j_l,s_l,\rho-\lambda} + \delta m_{j,P,F,j_l,s_l,\rho-\lambda} \, .
\end{equation}
Here $m_Q$ is the heavy quark mass; $\overline{\Lambda}_{F,j_l,s_l,\rho-\lambda} = \overline{\Lambda}_{|j_l-1/2|,P,F,j_l,s_l,\rho-\lambda} = \overline{\Lambda}_{j_l+1/2,P,F,j_l,s_l,\rho-\lambda}$ is the sum rule result obtained at the leading order; $\delta m_{j,P,F,j_l,s_l,\rho-\lambda}$ is the sum rule result obtained at the ${\mathcal O}(1/m_Q)$ order:
\begin{eqnarray}
\label{eq:masscorrection}
&& \delta m_{j,P,F,j_l,s_l,\rho-\lambda}
\\ \nonumber && = -\frac{1}{4m_{Q}}(K_{F,j_l,s_l,\rho-\lambda} + d_{M}C_{mag}\Sigma_{F,j_l,s_l,\rho-\lambda} ) \, ,
\end{eqnarray}
where $C_{mag} (m_{Q}/\mu) = [ \alpha_s(m_Q) / \alpha_s(\mu) ]^{3/\beta_0}$ with $\beta_0 = 11 - 2 n_f /3$, and the coefficient $d_{M} \equiv d_{j,j_{l}}$ is defined to be
\begin{eqnarray}
d_{j_{l}-1/2,j_{l}} &=& 2j_{l}+2\, ,
\\ \nonumber d_{j_{l}+1/2,j_{l}} &=& -2j_{l} \, ,
\end{eqnarray}
which is directly related to the baryon mass splitting within the same multiplet.

In Ref.~\cite{Chen:2016phw} we have systematically studied the $D$-wave heavy baryons of the $SU(3)$ flavor $\mathbf{\bar 3}_F$ using the method of QCD sum rules with HQET. In this paper we similarly study the $SU(3)$ flavor $\mathbf{6}_F$ ones, and calculate the analytical formulae for $\overline{\Lambda}_{F,j_l,s_l,\rho-\lambda}$, $K_{F,j_l,s_l,\rho-\lambda}$ and $\Sigma_{F,j_l,s_l,\rho-\lambda}$.
There are altogether seven heavy baryon multiplets of the $SU(3)$ flavor $\mathbf{6}_F$ as shown in Fig.~\ref{fig:dwave}, i.e., $[\mathbf{6}_F, 1, 1, \rho\rho]$, $[\mathbf{6}_F, 2, 1, \rho\rho]$, $[\mathbf{6}_F, 3, 1, \rho\rho]$, $[\mathbf{6}_F, 1, 1, \lambda\lambda]$, $[\mathbf{6}_F, 2, 1, \lambda\lambda]$, $[\mathbf{6}_F, 3, 1, \lambda\lambda]$ and $[\mathbf{6}_F, 2, 0, \rho\lambda]$. Hence, there can be as many as two $j^P = 1/2^+$ $D$-wave excited $\Sigma_c$ states, five $j^P = 3/2^+$ ones, five $j^P = 5/2^+$ ones, and two $j^P = 7/2^+$ ones. The numbers of excited $\Xi_c^\prime$ and $\Omega_c$ states are the same.
Recalling that the LHCb experiment observed as many as five excited $\Omega_c$ states~\cite{Aaij:2017nav}, some of these $D$-wave heavy baryons may be observed experimentally at the same time.
Theoretically, in the present study we can only use five baryon multiplets to perform sum rule analyses, because in Ref.~\cite{Chen:2016phw} we failed to construct the currents belonging to the other two multiplets $[\mathbf{6}_F, 2, 1, \rho\rho]$ and $[\mathbf{6}_F, 2, 1, \lambda\lambda]$.

As an example, we use the charmed baryon multiplet $[\Sigma_c(\mathbf{6}_F), 1, 1, \rho\rho]$ to perform QCD sum rule analyses. It contains two charmed baryons, $\Sigma_c(1/2^+)$ and $\Sigma_c(3/2^+)$, and the relevant interpolating field is
\begin{eqnarray}
&& J_{1/2,+,\Sigma_c,1,1,\rho\rho}(x)
\label{eq:current}
\\ \nonumber &=& \epsilon_{abc} \Big ( [\mathcal{D}^t_{\mu_1} \mathcal{D}^t_{\mu_2} u^{aT}(x)] \mathbb{C} \gamma^t_{\mu_3} d^b(x)
\\ \nonumber && ~~~~~~~~ - 2 [\mathcal{D}^t_{\mu_1} u^{aT}(x)] \mathbb{C} \gamma^t_{\mu_3} [\mathcal{D}^t_{\mu_2} d^b(x)]
\\ \nonumber && ~~~~~~~~~~~~~~~~ + u^{aT}(x) \mathbb{C} \gamma^t_{\mu_3} [\mathcal{D}^t_{\mu_1} \mathcal{D}^t_{\mu_2} d^b(x)] \Big )
\\ \nonumber && ~~~~~~~~
\times \Big ( g_t^{\mu_1 \mu_3} g_t^{\mu_2 \mu_4} + g_t^{\mu_2 \mu_3} g_t^{\mu_1 \mu_4} \Big ) \times \gamma^t_{\mu_4} \gamma_5 h_v^c(x) \, .
\end{eqnarray}
We can use this current to perform QCD sum rule analyses, and calculate $\overline{\Lambda}_{\Sigma_c,1,1,\rho\rho}$, $K_{\Sigma_c,1,1,\rho\rho}$ and $\Sigma_{\Sigma_c,1,1,\rho\rho}$:
\begin{eqnarray}
&& \Pi_{\Sigma_c,1,1,\rho\rho} = f_{\Sigma_c,1,1,\rho\rho}^{2} e^{-2 \bar \Lambda_{\Sigma_c,1,1,\rho\rho} / T}
\label{eq:ope}
\\ \nonumber &=& \int_{0}^{\omega_c} [\frac{11}{80640\pi^4}\omega^9 - \frac{19\langle g_s^2 GG \rangle}{3072\pi^4} \omega^5]e^{-\omega/T}d\omega \, ,
\\ && f_{\Sigma_c,1,1,\rho\rho}^{2} K_{\Sigma_c,1,1,\rho\rho} e^{-2 \bar \Lambda_{\Sigma_c,1,1,\rho\rho} / T}
\label{eq:Kc}
\\ \nonumber &=& \int_{0}^{\omega_c} [ - \frac{59}{2217600\pi^4}\omega^{11} + \frac{299\langle g_s^2 GG \rangle}{161280\pi^4} \omega^7]e^{-\omega/T}d\omega \, ,
\\ && f_{\Sigma_c,1,1,\rho\rho}^{2} \Sigma_{\Sigma_c,1,1,\rho\rho} e^{-2 \bar \Lambda_{\Sigma_c,1,1,\rho\rho} / T}
\label{eq:Sc}
\\ \nonumber &=& \int_{0}^{\omega_c} [\frac{37\langle g_s^2 GG \rangle}{322560\pi^4} \omega^7]e^{-\omega/T}d\omega \, .
\end{eqnarray}
Sum rules for $\Xi^\prime_c$ and $\Omega_c$ belonging to the same multiplet, $[\mathbf{6}_F, 1, 1, \rho\rho]$, as well as sum rules for the other four multiplets, $[\mathbf{6}_F, 3, 1, \rho\rho]$, $[\mathbf{6}_F, 1, 1, \lambda\lambda]$, $[\mathbf{6}_F, 3, 1, \lambda\lambda]$ and $[\mathbf{6}_F, 2, 0, \rho\lambda]$, are listed in Appendix.~\ref{app:sumrule}. In the next section we shall use these equations to perform numerical analyses.

\section{Numerical Analyses}
\label{sec:numerical}

We use the following values for various condensates and parameters to perform numerical analyses~\cite{pdg,Yang:1993bp,Hwang:1994vp,Narison:2002pw,Gimenez:2005nt,Jamin:2002ev,Ioffe:2002be,Ovchinnikov:1988gk,colangelo}:
%
\begin{eqnarray}
\nonumber && \langle \bar qq \rangle = - (0.24 \pm 0.01)^3 \mbox{ GeV}^3 \, ,
\\ \nonumber && \langle \bar ss \rangle = 0.8 \times \langle\bar qq \rangle \, ,
\\ \nonumber &&\langle g_s^2GG\rangle =(0.48\pm 0.14) \mbox{ GeV}^4\, ,
\\ \label{condensates} && \langle g_s \bar q \sigma G q \rangle = M_0^2 \times \langle \bar qq \rangle\, ,
\\ \nonumber && \langle g_s \bar s \sigma G s \rangle = M_0^2 \times \langle \bar ss \rangle\, ,
\\ \nonumber && m_s = 0.125 \mbox{ GeV} \, ,
\\ \nonumber && M_0^2= 0.8 \mbox{ GeV}^2\, .
\end{eqnarray}
Besides them, we also need the $charm$ quark mass, for which we use the PDG value $m_c = 1.275 \pm 0.025$ GeV~\cite{pdg} in the $\overline{\rm MS}$ scheme.

There are two free parameters in Eqs.~(\ref{eq:ope}--\ref{eq:Sc}): the Borel mass $T$ and the threshold value $\omega_c$. We follow Ref.~\cite{Chen:2016phw} and use three criteria to constrain them:
\begin{enumerate}

\item The first criterion requires that the high-order corrections should be less than 10\%:
%
\begin{equation}
\label{eq_convergence}
\mbox{Convergence (CVG)} \equiv |\frac{ \Pi^{\rm high-order}_{F,j_l,s_l,\rho-\lambda}(\infty, T) }{ \Pi_{F,j_l,s_l,\rho-\lambda}(\infty, T) }| \leq 10\% \, ,
\end{equation}
%
where $\Pi^{\rm high\mbox{-}order}_{F,j_l,s_l,\rho-\lambda}(\omega_c, T)$ denotes the high-order corrections, for example,
%
\begin{equation}
\Pi^{\rm high\mbox{-}order}_{\Sigma_c,1,1,\rho\rho}(\omega_c, T) = \int_{0}^{\omega_c} [ - \frac{19\langle g_s^2 GG \rangle}{3072\pi^4} \omega^5 ]e^{-\omega/T}d\omega \, .
\end{equation}
%

\item The second criterion requires that the pole contribution (PC) should be larger than 10\%:
%
\begin{equation}
\label{eq_pole}
\mbox{Pole Contribution (PC)} \equiv \frac{ \Pi_{F,j_l,s_l,\rho-\lambda}(\omega_c, T) }{ \Pi_{F,j_l,s_l,\rho-\lambda}( \infty , T) } \geq 10\% \, .
\end{equation}
%
We can use the first and second criteria together to arrive at an interval $T_{min}<T<T_{max}$ for a fixed threshold value $\omega_c$.

\item The third criterion requires that the dependence of $m_{j,P,F,j_l,s_l,\rho-\lambda}$ on the threshold value $\omega_c$ should be weak.

\end{enumerate}

\begin{figure*}[htbp]
\begin{center}
\scalebox{0.45}{\includegraphics{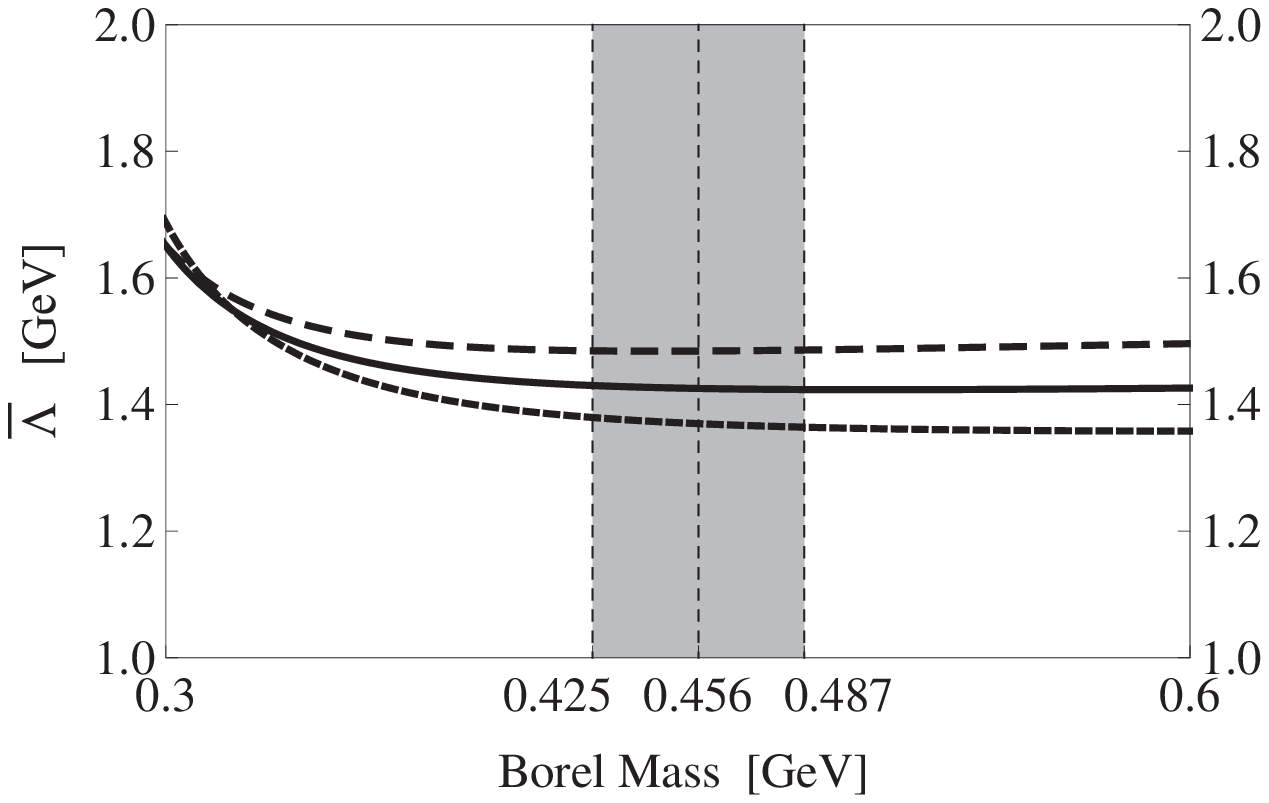}}
\scalebox{0.453}{\includegraphics{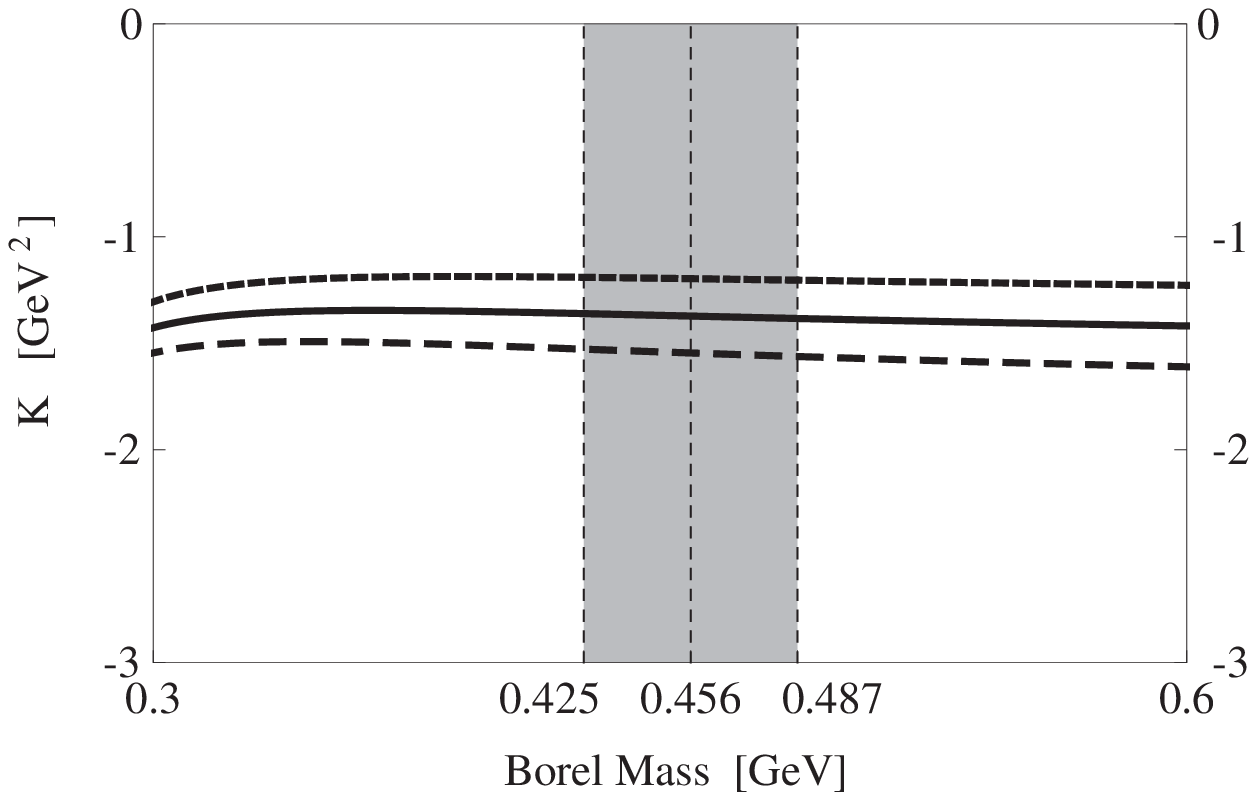}}
\scalebox{0.43}{\includegraphics{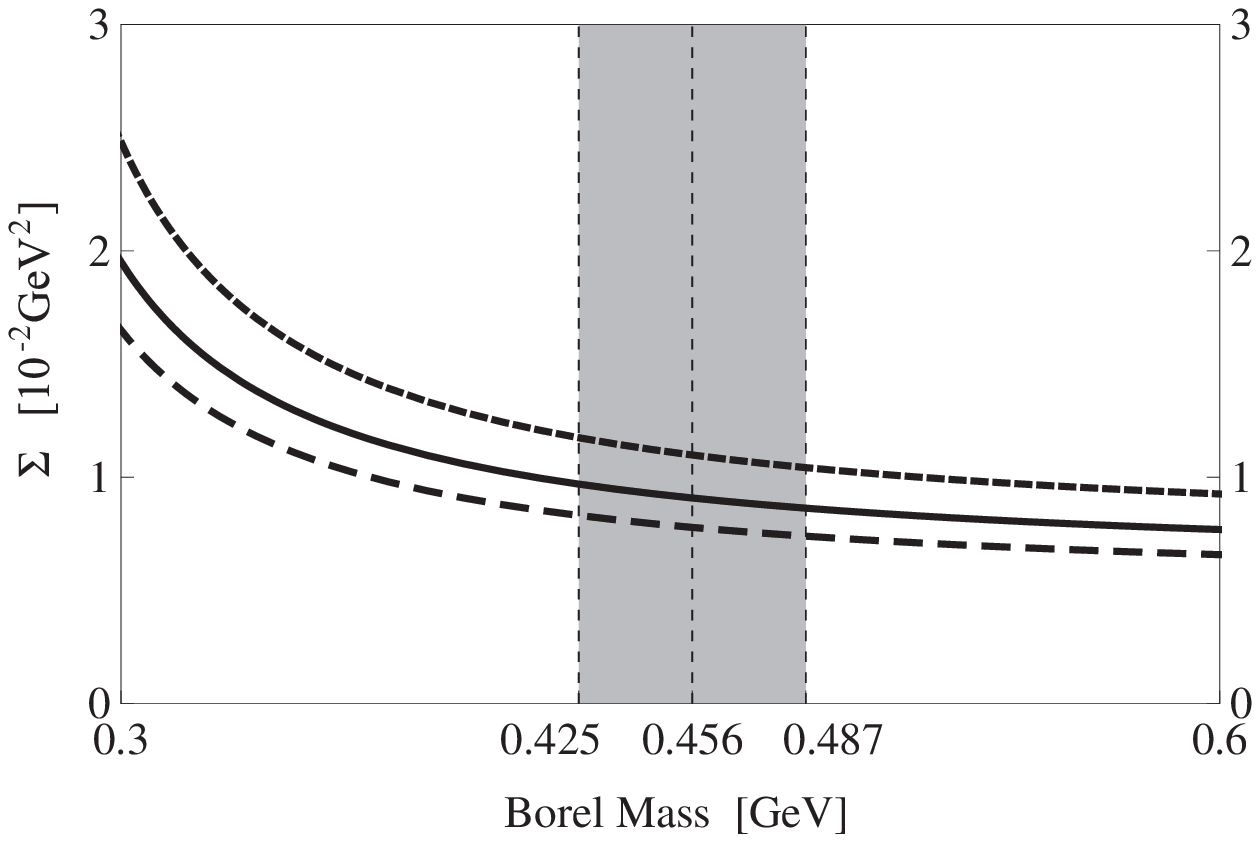}}
\caption{Variations of $\overline{\Lambda}_{\Sigma_c,1,1,\rho\rho}$ (left), $K_{\Sigma_c,1,1,\rho\rho}$ (middle), and $\Sigma_{\Sigma_c,1,1,\rho\rho}$ (right) with respect to the
Borel mass $T$, calculated using the charmed baryon doublet $[\Sigma_c(\mathbf{6}_F),1,1,\rho\rho]$.
The short-dashed, solid, and long-dashed curves are obtained by fixing $\omega_c = 3.0$, 3.2, and 3.4 GeV, respectively.}
\label{fig:piks}
\end{center}
\end{figure*}

\begin{figure*}[htbp]
\begin{center}
\begin{tabular}{c}
\scalebox{0.598}{\includegraphics{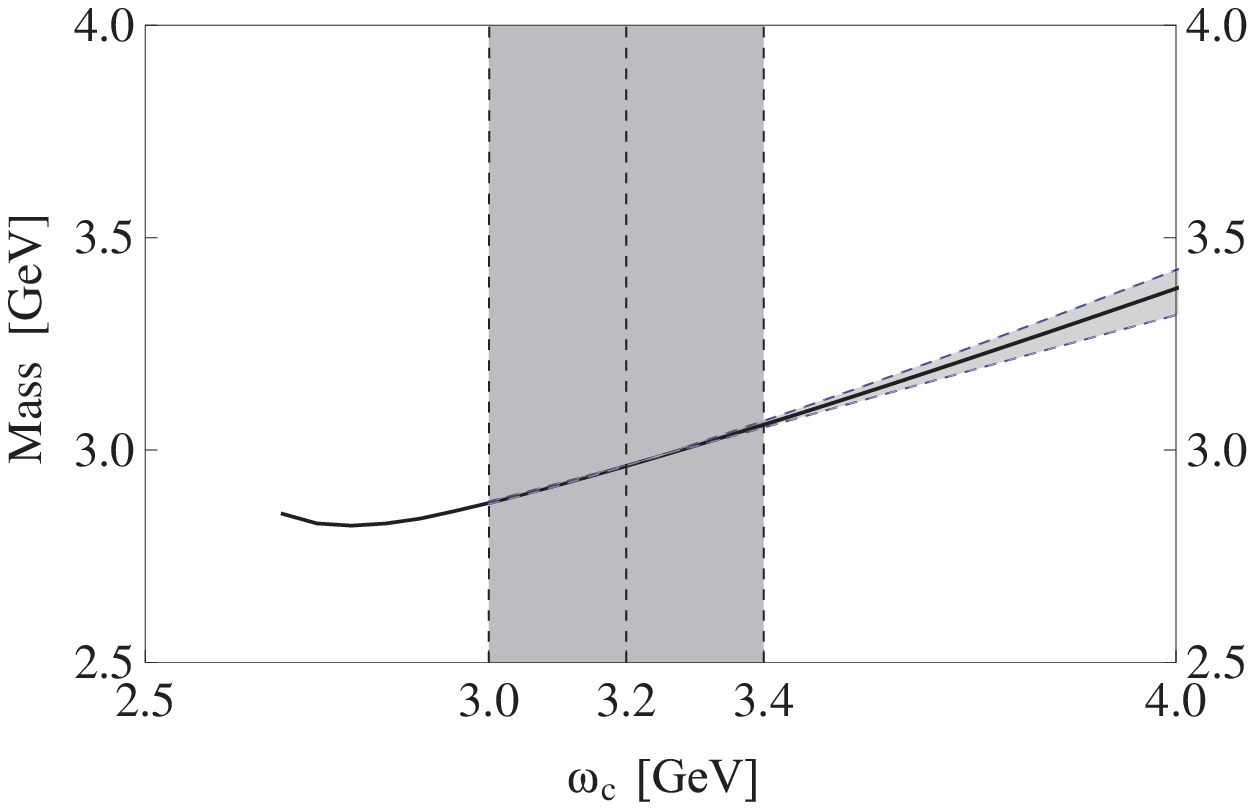}}
\scalebox{0.6}{\includegraphics{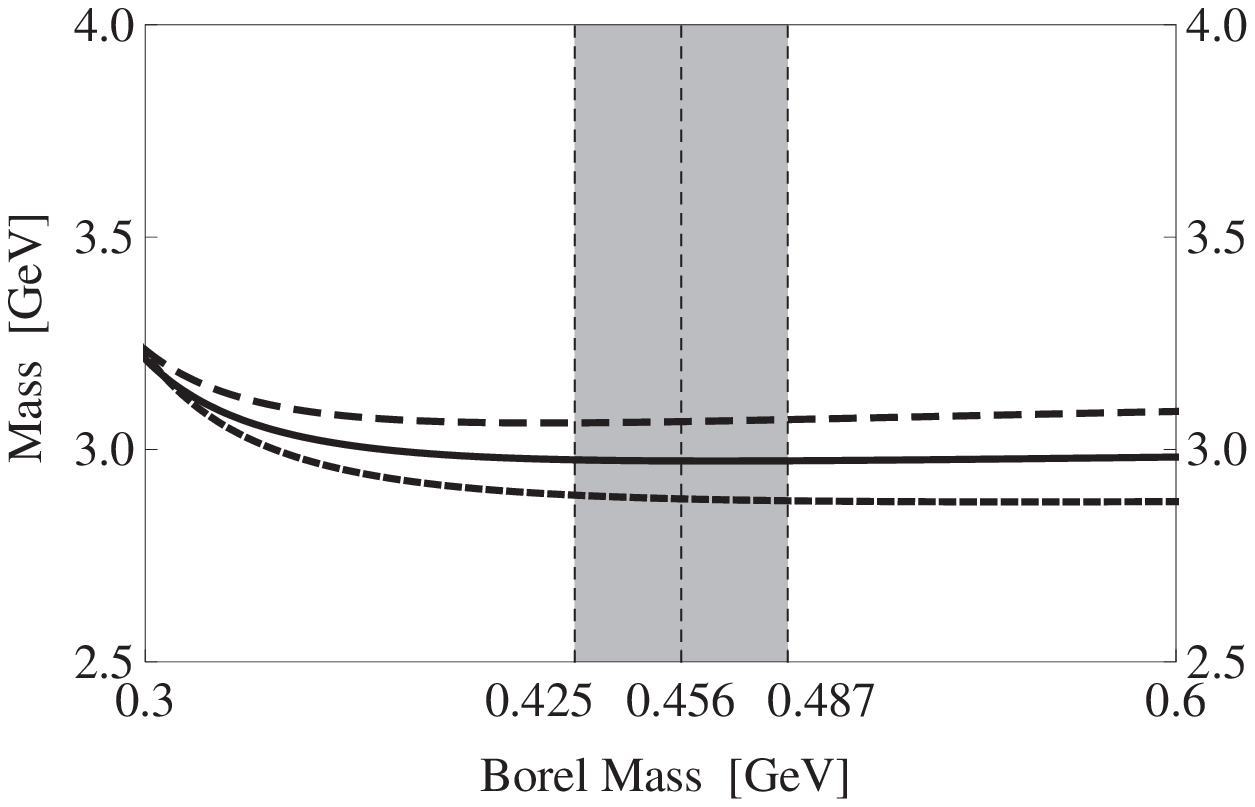}}
\end{tabular}
\caption{Variations of $m_{\Sigma_c(1/2^+)}$ with respect to the threshold value $\omega_c$ (left) and the Borel mass $T$ (right), calculated using the charmed baryon doublet $[\Sigma_c(\mathbf{6}_F),1,1,\rho\rho]$.
The shady band in the left panel is obtained by changing $T$ inside Borel windows. There exist non-vanishing working regions of $T$ as long as $\omega_c \geq 3.0$ GeV, while the results for $\omega_c < 3.0$ GeV are also shown, for which cases we choose the Borel mass $T$ when the PC defined in Eq.~(\ref{eq_pole}) is around 10\%.
In the right panel, the short-dashed, solid and long-dashed curves are obtained by setting $\omega_c = 3.0$, 3.2 and 3.4 GeV, respectively.
}
\label{fig:mass1120ud}
\end{center}
\end{figure*}

Still use the baryon multiplet $[\Sigma_c(\mathbf{6}_F), 1, 1, \rho\rho]$ as an example. Firstly, when we take $\omega_c = 3.2$ GeV, the Borel window can be evaluated to be $0.425$~GeV $< T < 0.487$~GeV, and the following numerical results can be obtained:
\begin{eqnarray}
\nonumber \bar \Lambda_{\Sigma_c,1,1,\rho\rho} &=& 1.425 \mbox{ GeV} \, , \,
\\ K_{\Sigma_c,1,1,\rho\rho} &=& -1.372 \mbox{ GeV}^2 \, , \,
\\ \nonumber \Sigma_{\Sigma_c,1,1,\rho\rho} &=& 0.0091 \mbox{ GeV}^{2} \, .
\end{eqnarray}
Their variations are shown in Fig.~\ref{fig:piks} with respect to the Borel mass $T$. Inserting them into Eqs.~(\ref{eq:mass}--\ref{eq:masscorrection}), we can further obtain
\begin{eqnarray}
\nonumber m_{\Sigma_c(1/2^+)} &=& 2.96 \mbox{ GeV} \, , \,
\\ m_{\Sigma_c(3/2^+)} &=& 2.97 \mbox{ GeV} \, , \,
\\ \nonumber \Delta m_{[\Sigma_c, 1, 1, \rho\rho]} &=& 11 \mbox{ MeV} \, ,
\end{eqnarray}
where $m_{\Sigma_c(1/2^+)}$ and $m_{\Sigma_c(3/2^+)}$ are the masses of the $\Sigma_c(1/2^+)$ and $\Sigma_c(3/2^+)$ belonging to the baryon multiplet $[\Sigma_c(\mathbf{6}_F), 1, 1, \rho\rho]$, and $\Delta m_{[\Sigma_c, 1, 1, \rho\rho]}$ is their mass splitting. We show the variation of $m_{\Sigma_b(1/2^+)}$ with respect to the Borel mass $T$ in the right panel of Fig.~\ref{fig:mass1120ud}.

Secondly, we change the threshold value $\omega_c$ and redo the above process. We show the variation of $m_{\Sigma_c(1/2^+)}$
with respect to the threshold value $\omega_c$ in the left panel of Fig.~\ref{fig:mass1120ud}. There are non-vanishing Borel windows as long as $\omega_c \geq 3.0$ GeV, and
the $\omega_c$ dependence is weak and acceptable in the region $3.0$ GeV$<\omega_c<3.4$ GeV. The results for $\omega_c \leq 3.0$ GeV are also shown, for which cases we choose the Borel mass $T$ when the PC defined in Eq.~(\ref{eq_pole}) is around 10\%.

Finally, we choose our working regions to be $3.0$ GeV$<\omega_c<3.4$ GeV and $0.425$ GeV $< T < 0.487$ GeV, and obtain
the following numerical results for the baryon multiplet $[\Sigma_c(\mathbf{6}_F), 1, 1, \rho\rho]$:
\begin{eqnarray}
\nonumber m_{\Sigma_c(1/2^+)} &=& 2.96^{+0.17}_{-0.12} \mbox{ GeV} \, , \,
\\ m_{\Sigma_c(3/2^+)} &=& 2.97^{+0.18}_{-0.13} \mbox{ GeV} \, , \,
\\ \nonumber \Delta m_{[\Sigma_c, 1, 1, \rho\rho]} &=& 11^{+17}_{-9} \mbox{ MeV} \, ,
\end{eqnarray}
where the central values correspond to $\omega_c = 3.2$ GeV and $T=0.456$ GeV,
and the uncertainties come from the threshold value $\omega_c$, the Borel mass $T$, the charm quark mass $m_c$, and various quark and gluon condensates.
We note that there are large theoretical uncertainties in our mass predictions, but the mass splitting within
the same doublet is produced quite well with much less theoretical uncertainty, because it does not depend much on the charm quark mass~\cite{Chen:2016phw,Liu:2007fg,Chen:2015kpa,Mao:2015gya}.

To study the charmed baryon multiplets $[\Xi_c^\prime(\mathbf{6}_F),1,1,\rho\rho]$ and $[\Omega_c(\mathbf{6}_F),1,1,\rho\rho]$, we fine-tune the threshold value $\omega_c$ to be
\begin{eqnarray}
\omega_c([\Xi_c^\prime,1,1,\rho\rho]) &\approx& 3.7~{\rm GeV} \, ,
\\ \nonumber \omega_c([\Omega_c,1,1,\rho\rho]) &\approx& 4.2~{\rm GeV} \, ,
\end{eqnarray}
so that
\begin{eqnarray}
&& \omega_c([\Omega_c,1,1,\rho\rho]) - \omega_c([\Xi_c^\prime,1,1,\rho\rho]) \approx
\\ \nonumber && \omega_c([\Xi_c^\prime,1,1,\rho\rho]) - \omega_c([\Sigma_c,1,1,\rho\rho]) \approx 0.5~{\rm GeV} \, ,
\end{eqnarray}
which value is the same as those used in our previous studies on excited heavy baryons~\cite{Chen:2016phw,Chen:2015kpa,Mao:2015gya}.

\begin{figure*}[htbp]
\begin{center}
\begin{tabular}{c}
\scalebox{0.6}{\includegraphics{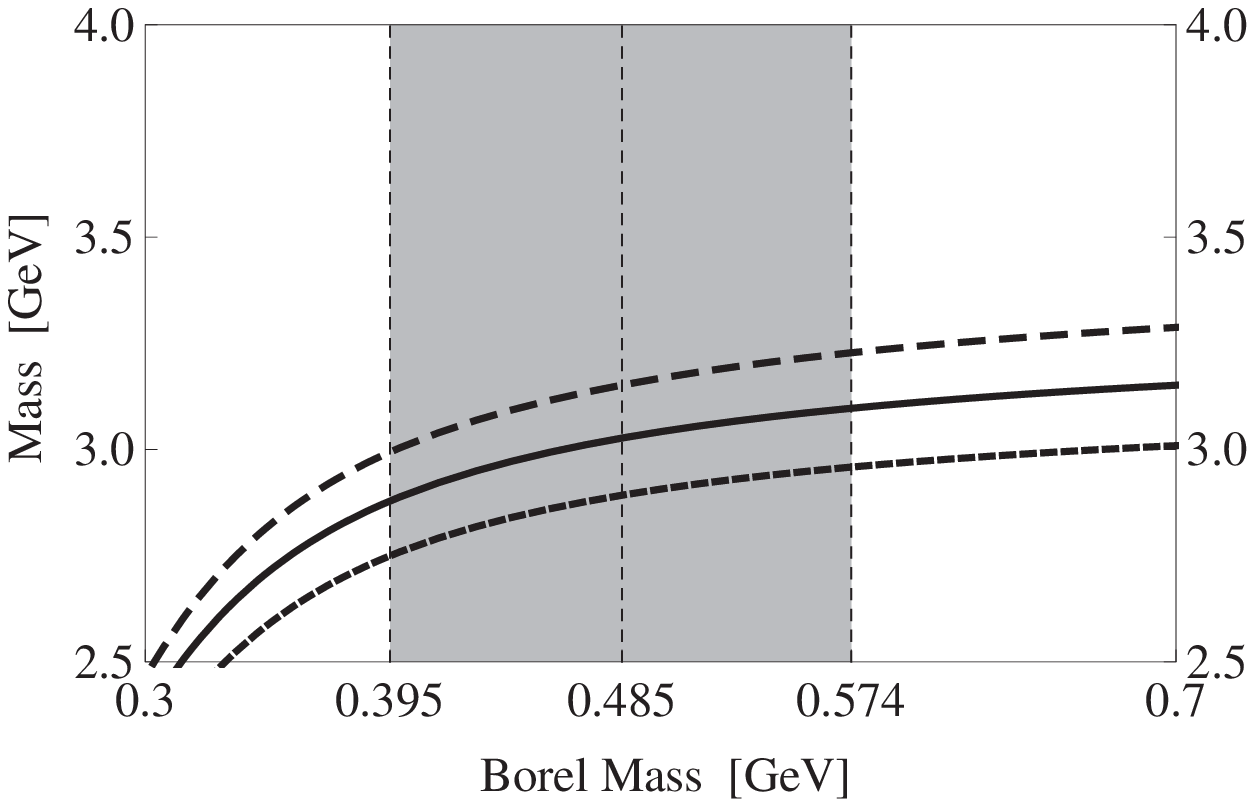}}
\scalebox{0.6}{\includegraphics{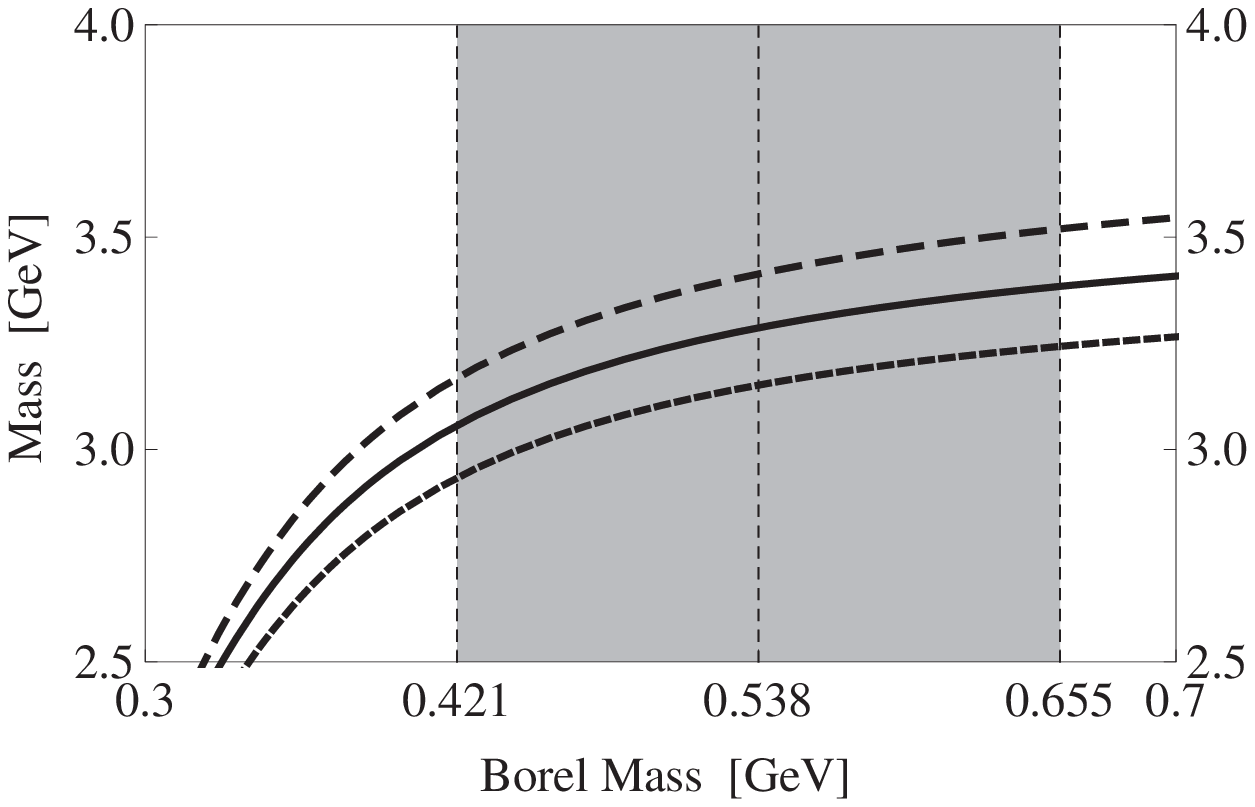}}
\end{tabular}
\caption{Variations of $m_{\Xi_c^\prime(1/2^+)}$ (left) and $m_{\Omega_c(1/2^+)}$ (right) with respect to the Borel mass $T$, calculated using the charmed baryon multiplets $[\Xi_c^\prime(\mathbf{6}_F),1,1,\rho\rho]$ and $[\Omega_c(\mathbf{6}_F),1,1,\rho\rho]$.
In the left figure, the short-dashed, solid and long-dashed curves are obtained by setting $\omega_c = 3.5$, 3.7 and 3.9 GeV, respectively.
In the left figure, the short-dashed, solid and long-dashed curves are obtained by setting $\omega_c = 4.0$, 4.2 and 4.4 GeV, respectively.
}
\label{fig:mass1120other}
\end{center}
\end{figure*}

After fixing the threshold value $\omega_c([\Xi_c^\prime,1,1,\rho\rho])$ to be around 3.7 GeV, we can evaluate our working regions to be $3.5$ GeV$<\omega_c<3.9$ GeV and $0.395$ GeV $< T < 0.574$ GeV, and obtain
the following numerical results for the charmed baryon multiplet $[\Xi_c^\prime(\mathbf{6}_F), 1, 1, \rho\rho]$:
\begin{eqnarray}
\nonumber m_{\Xi_c^\prime(1/2^+)} &=& 3.02^{+0.15}_{-0.21} \mbox{ GeV} \, , \,
\\ m_{\Xi_c^\prime(3/2^+)} &=& 3.03^{+0.15}_{-0.21} \mbox{ GeV} \, , \,
\\ \nonumber \Delta m_{[\Xi_c^\prime, 1, 1, \rho\rho]} &=& 7^{+8}_{-6} \mbox{ MeV} \, ,
\end{eqnarray}
where the central values correspond to $\omega_c = 3.7$ GeV and $T=0.485$ GeV.
We show the variation of $m_{\Xi_c^\prime(1/2^+)}$ with respect to the Borel mass $T$ in the left panel of Fig.~\ref{fig:mass1120other}, where these curves are stable inside the Borel window $0.395$ GeV $< T < 0.574$ GeV.
The masses of $\Xi_c^\prime(1/2^+)$ and $\Xi_c^\prime(3/2^+)$ are both consistent with the mass of the $\Xi_c(3123)$ observed by the BaBar Collaboration~\cite{Aubert:2007dt}:
\begin{eqnarray}
m^{\rm exp}_{\Xi_c(3123)} = 3122.9 \pm 1.3 \pm 0.3 \mbox{ MeV} \, ,
\end{eqnarray}
which supports it to be a $D$-wave $\Xi_c^\prime$ state of $J^P=1/2^+$ or $3/2^+$.

Similarly, after fixing $\omega_c([\Omega_c,1,1,\rho\rho])$ to be around 4.2 GeV, we can evaluate our working regions to be $4.0$ GeV$<\omega_c<4.4$ GeV and $0.421$ GeV $< T < 0.655$ GeV, and obtain
the following numerical results for the charmed baryon multiplet $[\Omega_c(\mathbf{6}_F), 1, 1, \rho\rho]$:
\begin{eqnarray}
\nonumber m_{\Omega_c(1/2^+)} &=& 3.29^{+0.17}_{-0.25} \mbox{ GeV} \, , \,
\\ m_{\Omega_c(3/2^+)} &=& 3.29^{+0.16}_{-0.25} \mbox{ GeV} \, , \,
\\ \nonumber \Delta m_{[\Omega_c, 1, 1, \rho\rho]} &=& 5^{+5}_{-4} \mbox{ MeV} \, ,
\end{eqnarray}
where the central values correspond to $\omega_c = 4.2$ GeV and $T=0.538$ GeV.
We show the variation of $m_{\Omega_c(1/2^+)}$ with respect to the Borel mass $T$ in the right panel of Fig.~\ref{fig:mass1120other}, where these curves are stable inside the Borel window $0.421$ GeV $< T < 0.655$ GeV.

\begin{table*}[hbt]
\begin{center}
\renewcommand{\arraystretch}{1.6}
\caption{Masses of the $D$-wave charmed baryons of the $SU(3)$ flavor $\mathbf{6}_F$, obtained using the charmed baryon multiplets $[\mathbf{6}_F, 1, 1, \rho\rho]$, $[\mathbf{6}_F, 3, 1, \lambda\lambda]$ and $[\mathbf{6}_F, 2, 0, \rho\lambda]$.
For the charmed baryon multiplet $[\mathbf{6}_F, 3, 1, \lambda\lambda]$ containing $\Sigma_c(5/2^+,7/2^+)$, $\Xi_c^\prime(5/2^+,7/2^+)$, and $\Omega_c(5/2^+,7/2^+)$, we can only evaluate their average masses, i.e., ${1\over14} ( 6 m_{\Sigma_c({5/2}^+)} + 8 m_{\Sigma_c({7/2}^+)} )$, ${1\over14} ( 6 m_{\Xi_c^\prime({5/2}^+)} + 8 m_{\Xi_c^\prime({7/2}^+)} )$ and ${1\over14} ( 6 m_{\Omega_c({5/2}^+)} + 8 m_{\Omega_c({7/2}^+)} )$, as discussed in Ref.~\cite{Chen:2016phw}.
For the baryon multiplet $[\mathbf{6}_F, 2, 0, \rho\lambda]$ containing $\Sigma_c(3/2^+,5/2^+)$, $\Xi_c^\prime(3/2^+,5/2^+)$, and $\Omega_c(3/2^+,5/2^+)$, the mass differences among $\Sigma_c$, $\Xi_c^\prime$, and $\Omega_c$ seem a bit large, so we do not use them to draw conclusions.
}
\begin{tabular}{c | c | c | c | c c | c c | c | c | c}
\hline\hline
\multirow{2}{*}{Multiplets} & \multirow{2}{*}{B} & $\omega_c$ & Working region & $\overline{\Lambda}$ & $f$ & $K$ & $\Sigma$ & Baryons & Mass & Difference
\\ & & (GeV) & (GeV) & (GeV) & (GeV$^{5}$) & (GeV$^2$) & (GeV$^2$) & ($j^P$) & (GeV) & (MeV)
\\ \hline\hline
\multirow{6}{*}{$[\mathbf{6}_F, 1, 1, \rho\rho]$} & \multirow{2}{*}{$\Sigma_c$} & \multirow{2}{*}{3.2} & \multirow{2}{*}{$0.425< T < 0.487$} & \multirow{2}{*}{$1.425$} & \multirow{2}{*}{$0.079$} & \multirow{2}{*}{$-1.372$} & \multirow{2}{*}{$0.0091$} & $\Sigma_c(1/2^+)$ & $2.96^{+0.17}_{-0.12}$ & \multirow{2}{*}{$11^{+17}_{-9}$}
\\ \cline{9-10}
& & & & & & & & $\Sigma_c(3/2^+)$ & $2.97^{+0.18}_{-0.13}$ &
\\ \cline{2-11}
& \multirow{2}{*}{$\Xi_c^\prime$} & \multirow{2}{*}{3.7} & \multirow{2}{*}{$0.395< T < 0.574$} & \multirow{2}{*}{$1.561$} & \multirow{2}{*}{$0.146$} & \multirow{2}{*}{$-0.961$} & \multirow{2}{*}{$0.0057$}
& $\Xi_c^\prime(1/2^+)$ & $3.02^{+0.15}_{-0.21}$ & \multirow{2}{*}{$7^{+8}_{-6}$}
\\ \cline{9-10}
& & & & & & & & $\Xi_c^\prime(3/2^+)$ & $3.03^{+0.15}_{-0.21}$ &
\\ \cline{2-11}
& \multirow{2}{*}{$\Omega_c$} & \multirow{2}{*}{4.2} & \multirow{2}{*}{$0.421< T < 0.655$} & \multirow{2}{*}{$1.761$} & \multirow{2}{*}{$0.274$} & \multirow{2}{*}{$-1.302$} & \multirow{2}{*}{$0.0042$}
& $\Omega_c(1/2^+)$ & $3.29^{+0.17}_{-0.25}$ & \multirow{2}{*}{$5^{+5}_{-4}$}
\\ \cline{9-10}
& & & & & & & & $\Omega_c(3/2^+)$ & $3.29^{+0.16}_{-0.25}$ &
\\ \hline
\multirow{6}{*}{$[\mathbf{6}_F, 3, 1, \lambda\lambda]$} & \multirow{2}{*}{$\Sigma_c$} & \multirow{2}{*}{3.1} & \multirow{2}{*}{$0.445< T < 0.459$} & \multirow{2}{*}{$1.432$} & \multirow{2}{*}{$0.014$} & \multirow{2}{*}{$-3.139$} & \multirow{2}{*}{--}
& $\Sigma_c(5/2^+)$ & \multirow{2}{*}{$3.32^{+0.64}_{-0.24}$} & \multirow{2}{*}{--}
\\ \cline{9-9}
& & & & & & & & $\Sigma_c(7/2^+)$ & &
\\ \cline{2-11}
& \multirow{2}{*}{$\Xi_c^\prime$} & \multirow{2}{*}{3.3} & \multirow{2}{*}{$0.458< T < 0.485$} & \multirow{2}{*}{$1.509$} & \multirow{2}{*}{$0.018$} & \multirow{2}{*}{$-3.064$} & \multirow{2}{*}{--}
& $\Xi_c^\prime(5/2^+)$ & \multirow{2}{*}{$3.39^{+0.40}_{-0.20}$} & \multirow{2}{*}{--}
\\ \cline{9-9}
& & & & & & & & $\Xi_c^\prime(7/2^+)$ & &
\\ \cline{2-11}
& \multirow{2}{*}{$\Omega_c$} & \multirow{2}{*}{3.5} & \multirow{2}{*}{$0.490< T < 0.509$} & \multirow{2}{*}{$1.612$} & \multirow{2}{*}{$0.024$} & \multirow{2}{*}{$-3.085$} & \multirow{2}{*}{--}
& $\Omega_c(5/2^+)$ & \multirow{2}{*}{$3.49^{+0.30}_{-0.19}$} & \multirow{2}{*}{--}
\\ \cline{9-9}
& & & & & & & & $\Omega_c(7/2^+)$ & &
\\ \hline
\multirow{6}{*}{$[\mathbf{6}_F, 2, 0, \rho\lambda]$} & \multirow{2}{*}{$\Sigma_c$} & \multirow{2}{*}{3.0} & \multirow{2}{*}{$0.398< T < 0.457$} & \multirow{2}{*}{$1.336$} & \multirow{2}{*}{$0.022$} & \multirow{2}{*}{$-2.156$} & \multirow{2}{*}{$0.0082$} & $\Sigma_c(3/2^+)$ & $3.02^{+0.26}_{-0.16}$ & \multirow{2}{*}{$16^{+26}_{-14}$}
\\ \cline{9-10}
& & & & & & & & $\Sigma_c(5/2^+)$ & $3.04^{+0.28}_{-0.17}$ &
\\ \cline{2-11}
& \multirow{2}{*}{$\Xi_c^\prime$} & \multirow{2}{*}{3.7} & \multirow{2}{*}{$0.505< T < 0.543$} & \multirow{2}{*}{$1.793$} & \multirow{2}{*}{$0.076$} & \multirow{2}{*}{$-2.214$} & \multirow{2}{*}{$0.0055$}
& $\Xi_c^\prime(3/2^+)$ & $3.50^{+0.15}_{-0.13}$ & \multirow{2}{*}{$11^{+12}_{-9}$}
\\ \cline{9-10}
& & & & & & & & $\Xi_c^\prime(5/2^+)$ & $3.51^{+0.15}_{-0.13}$ &
\\ \cline{2-11}
& \multirow{2}{*}{$\Omega_c$} & \multirow{2}{*}{4.0} & \multirow{2}{*}{$0.561< T < 0.575$} & \multirow{2}{*}{$1.986$} & \multirow{2}{*}{$0.115$} & \multirow{2}{*}{$-2.121$} & \multirow{2}{*}{$0.0046$}
& $\Omega_c(3/2^+)$ & $3.67^{+0.14}_{-0.14}$ & \multirow{2}{*}{$9^{+9}_{-8}$}
\\ \cline{9-10}
& & & & & & & & $\Omega_c(5/2^+)$ & $3.68^{+0.15}_{-0.14}$ &
\\ \hline \hline
\end{tabular}
\label{tab:results}
\end{center}
\end{table*}

We list the above results in Table~\ref{tab:results}. Following the same procedures, we study the charmed baryon multiplets, $[\mathbf{6}_F, 3, 1, \lambda\lambda]$, $[\mathbf{6}_F, 2, 0, \rho\lambda]$, $[\mathbf{6}_F, 3, 1, \rho\rho]$, $[\mathbf{6}_F, 1, 1, \lambda\lambda]$, and obtain:
\begin{enumerate}

\item The baryon multiplet $[\mathbf{6}_F, 3, 1, \lambda\lambda]$ contains $\Sigma_c(5/2^+,7/2^+)$, $\Xi_c^\prime(5/2^+,7/2^+)$, and $\Omega_c(5/2^+,7/2^+)$. We use them to perform sum rule analyses, and obtain:
\begin{eqnarray}
\nonumber m_{[\Sigma_c, 3, 1, \lambda\lambda]} &\sim& 3.32^{+0.64}_{-0.24} \mbox{ GeV} \, ,
\\ m_{[\Xi^\prime_c, 3, 1, \lambda\lambda]} &\sim& 3.39^{+0.40}_{-0.20} \mbox{ GeV} \, ,
\\ \nonumber m_{[\Omega_c, 3, 1, \lambda\lambda]} &\sim& 3.49^{+0.30}_{-0.19} \mbox{ GeV} \, .
\end{eqnarray}
These values are also listed in Table~\ref{tab:results}, and their variations are shown in Fig.~\ref{fig:mass3102} with respect to the threshold value $\omega_c$.
The masses of $\Xi_c^\prime(5/2^+,7/2^+)$ are not far from the mass of the $\Xi_c(3123)$~\cite{Aubert:2007dt}, suggesting that the $\Xi_c(3123)$ may also be interpreted as a $D$-wave $\Xi_c^\prime$ state of $J^P=5/2^+$ or $7/2^+$.

We note that we can only evaluate their average masses ${1\over14} ( 6 m_{\Sigma_c({5/2}^+)} + 8 m_{\Sigma_c({7/2}^+)} )$, ${1\over14} ( 6 m_{\Xi_c^\prime({5/2}^+)} + 8 m_{\Xi_c^\prime({7/2}^+)} )$ and ${1\over14} ( 6 m_{\Omega_c({5/2}^+)} + 8 m_{\Omega_c({7/2}^+)} )$, because their mass splittings related to the chromomagnetic interaction ($\Sigma_{\mathbf{6}_F, 3, 1, \lambda\lambda}$) are difficult to be evaluated. More discussions can be found in Ref.~\cite{Chen:2016phw}.

\begin{figure*}[hbtp]
\begin{center}
\begin{tabular}{c}
\scalebox{0.45}{\includegraphics{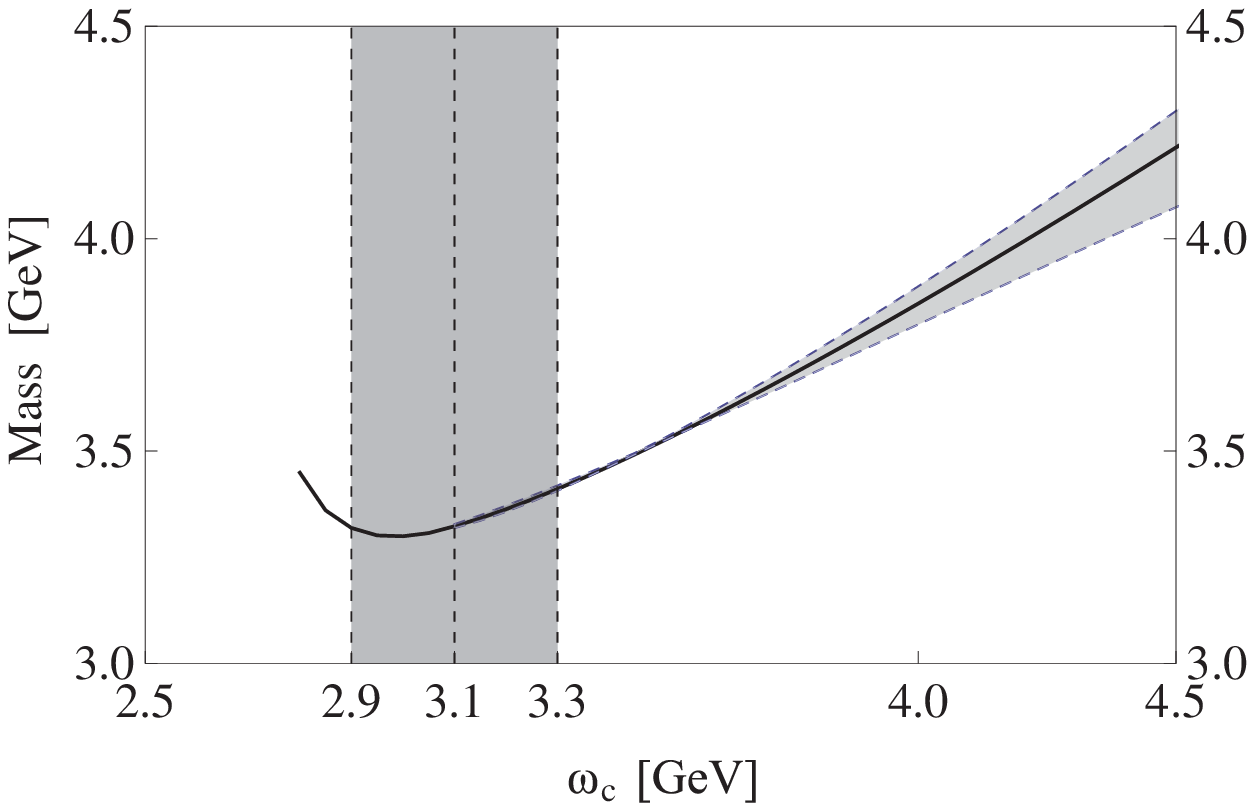}}
\scalebox{0.45}{\includegraphics{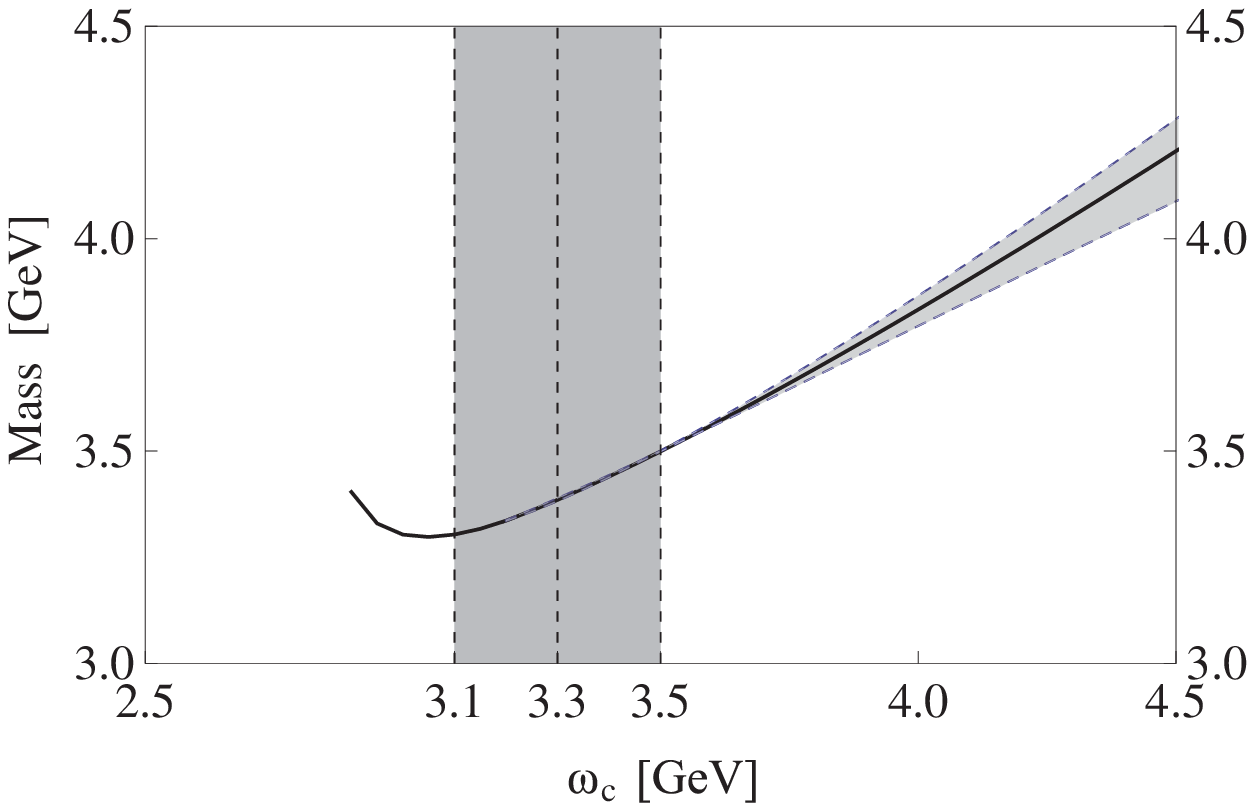}}
\scalebox{0.45}{\includegraphics{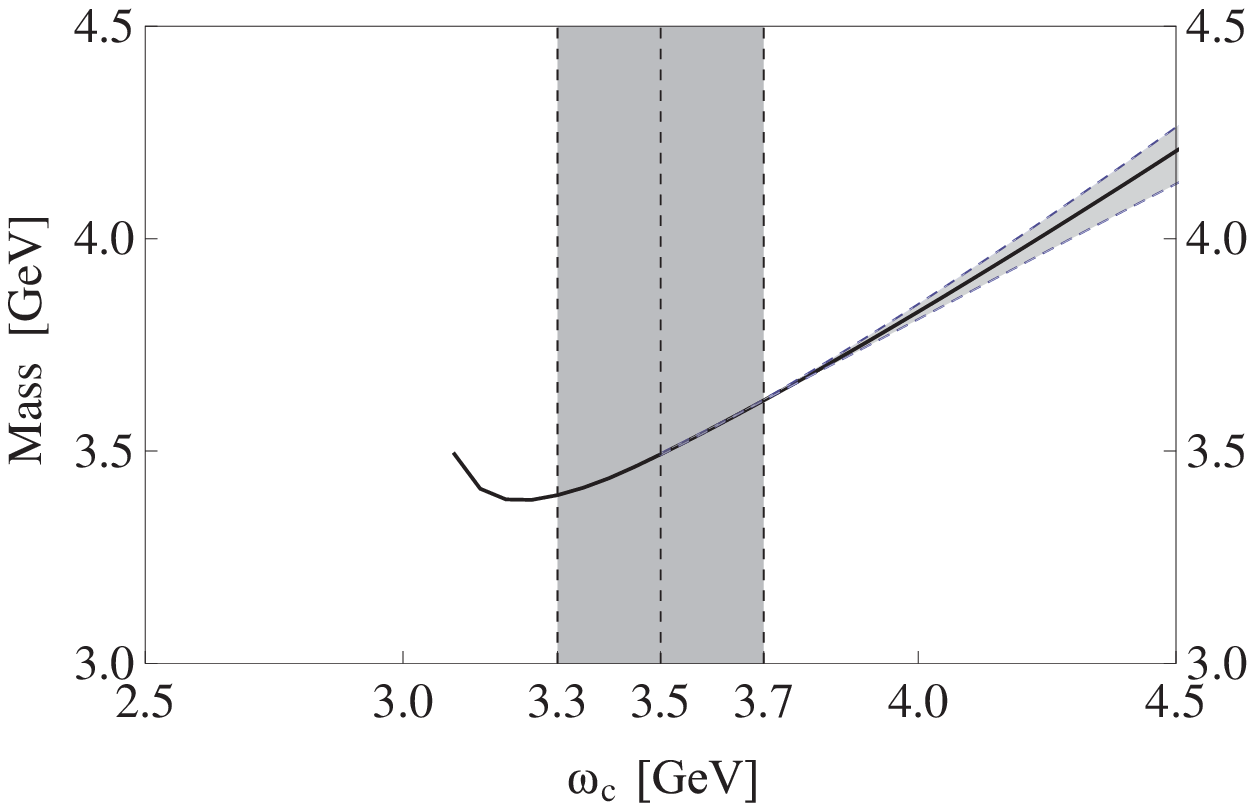}}
\end{tabular}
\caption{Variations of $m_{[\Sigma_c, 3, 1, \lambda\lambda]}$ (left), $m_{[\Xi^\prime_c, 3, 1, \lambda\lambda]}$ (middle) and $m_{[\Omega_c, 3, 1, \lambda\lambda]}$ (right) with respect to the threshold value $\omega_c$, calculated using the charmed baryon doublet $[\mathbf{6}_F, 3, 1, \lambda\lambda]$. The shady bands in these figures are obtained by changing $T$ inside Borel windows.
There exist Borel windows as long as $\omega_c([\Sigma_c, 3, 1, \lambda\lambda]) \geq 3.1$ GeV (left), $\omega_c([\Xi^\prime_c, 3, 1, \lambda\lambda]) \geq 3.2$ GeV (left) and $\omega_c([\Omega_c, 3, 1, \lambda\lambda]) \geq 3.5$ GeV (right).
Accordingly, we choose $\omega_c$ to be around 3.1, 3.3, and 3.5 GeV in the left, middle and right panels, respectively.
}
\label{fig:mass3102}
\end{center}
\end{figure*}

\item The baryon multiplet $[\mathbf{6}_F, 2, 0, \rho\lambda]$ contains $\Sigma_c(3/2^+,5/2^+)$, $\Xi_c^\prime(3/2^+,5/2^+)$, and $\Omega_c(3/2^+,5/2^+)$. We use them to perform sum rule analyses, and obtain:
\begin{eqnarray}
\nonumber m_{\Sigma_c({3/2}^+)} &=& 3.02^{+0.26}_{-0.16} \mbox{ GeV} \, ,
\\ \nonumber m_{\Sigma_c({5/2}^+)} &=& 3.04^{+0.28}_{-0.17} \mbox{ GeV} \, ,
\\ \nonumber \Delta m_{[\Sigma_c, 2, 0, \rho\lambda]} &=& 16^{+26}_{-14} \mbox{ MeV} \, ,
\\ \nonumber m_{\Xi_c^\prime({3/2}^+)} &=& 3.50^{+0.15}_{-0.13} \mbox{ GeV} \, ,
\\ m_{\Xi_c^\prime({5/2}^+)} &=& 3.51^{+0.15}_{-0.13} \mbox{ GeV} \, ,
\\ \nonumber \Delta m_{[\Xi_c^\prime, 2, 0, \rho\lambda]} &=& 11^{+12}_{-9} \mbox{ MeV} \, ,
\\ \nonumber m_{\Omega_c({3/2}^+)} &=& 3.67^{+0.14}_{-0.14} \mbox{ GeV} \, ,
\\ \nonumber m_{\Omega_c({5/2}^+)} &=& 3.68^{+0.15}_{-0.14} \mbox{ GeV} \, ,
\\ \nonumber \Delta m_{[\Omega_c, 2, 0, \rho\lambda]} &=& 9^{+9}_{-8} \mbox{ MeV} \, .
\end{eqnarray}
These values are also listed in Table~\ref{tab:results}, and variations of $m_{\Sigma_c(3/2^+)}$, $m_{\Xi_c^\prime(3/2^+)}$ and $m_{\Omega_c(3/2^+)}$ are shown in Fig.~\ref{fig:mass2011} with respect to the threshold value $\omega_c$.
However, the mass differences among $\Sigma_c(3/2^+,5/2^+)$, $\Xi_c^\prime(3/2^+,5/2^+)$ and $\Omega_c(3/2^+,5/2^+)$ seem a bit large, and we shall not use them to draw conclusions.

\begin{figure*}[hbtp]
\begin{center}
\begin{tabular}{c}
\scalebox{0.45}{\includegraphics{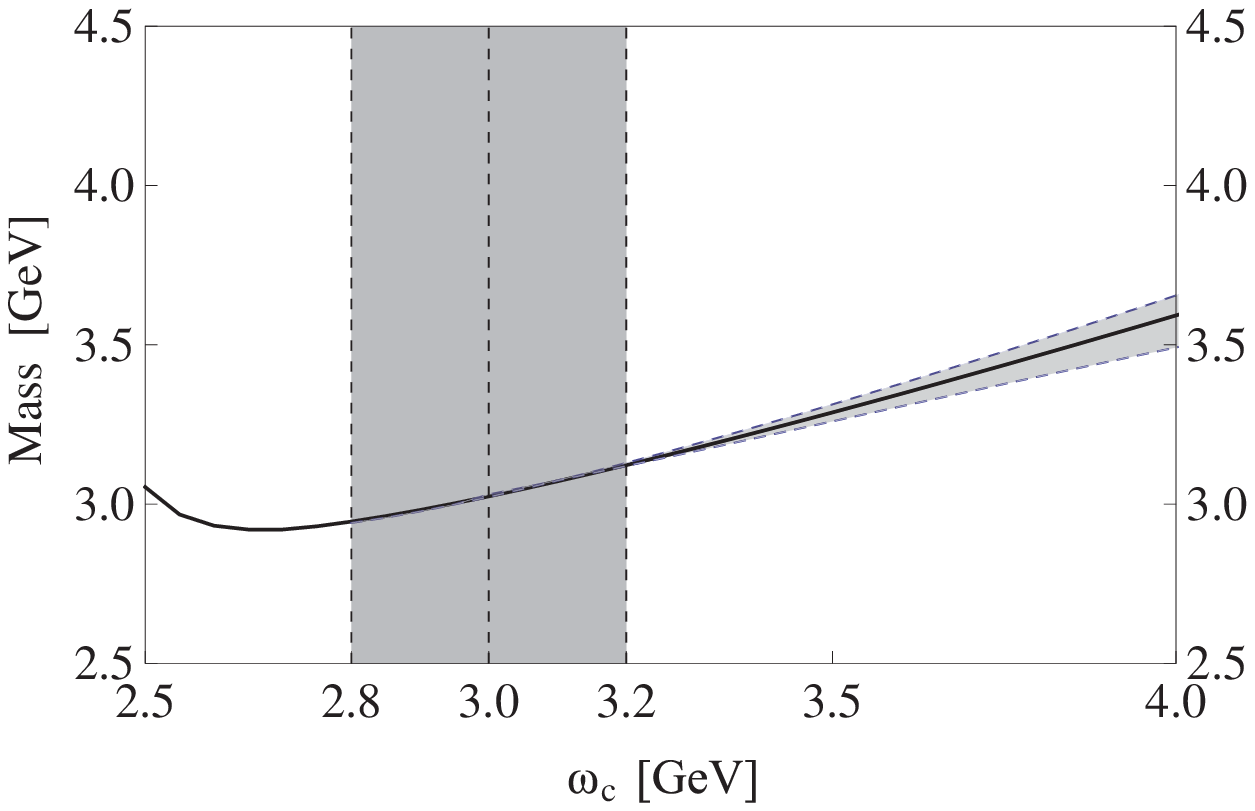}}
\scalebox{0.45}{\includegraphics{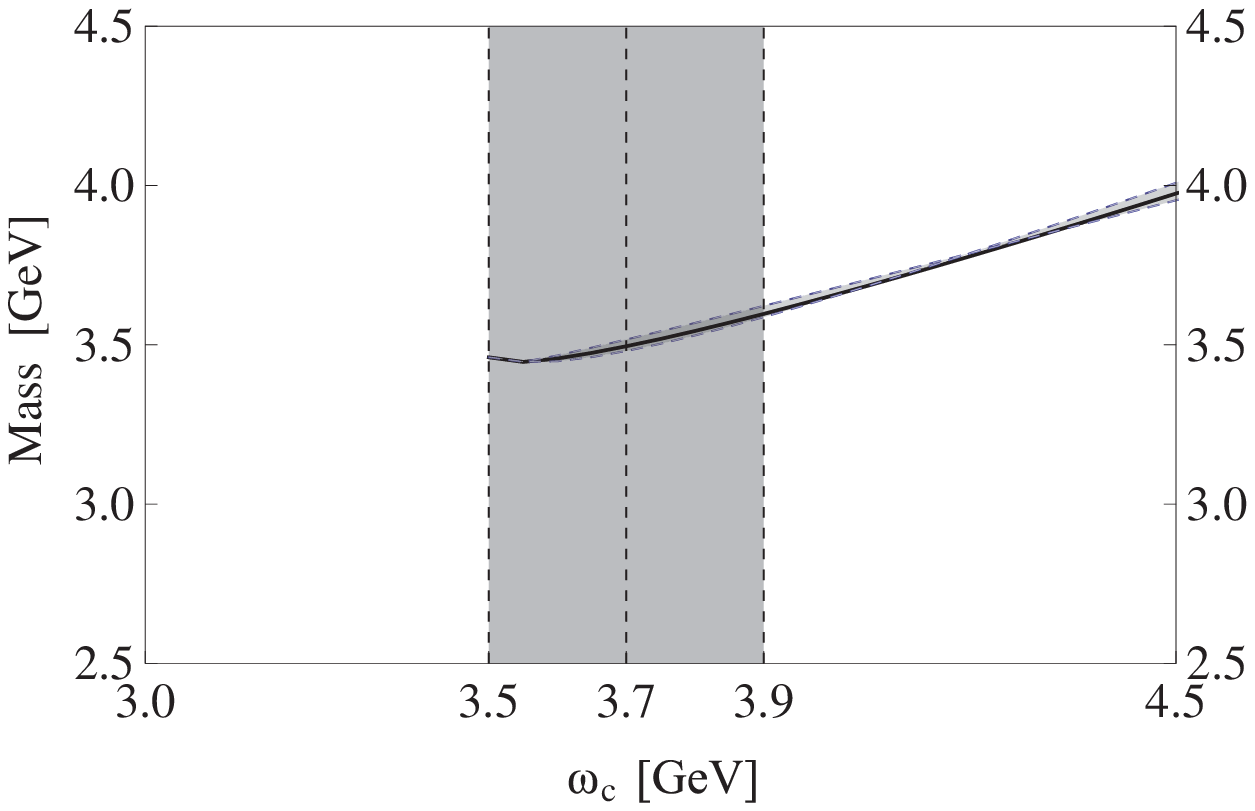}}
\scalebox{0.45}{\includegraphics{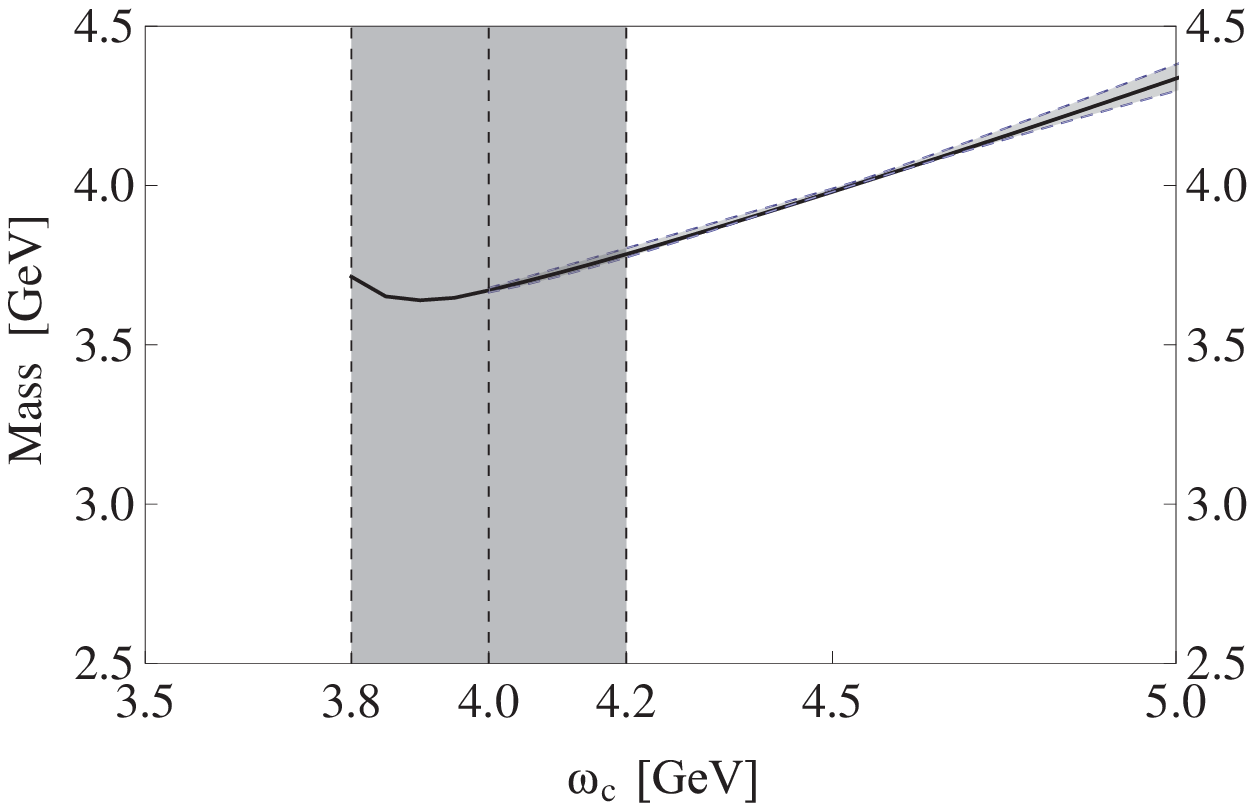}}
\end{tabular}
\caption{Variations of $m_{\Sigma_c(3/2^+)}$ (left), $m_{\Xi_c^\prime(3/2^+)}$ (middle) and $m_{\Omega_c(5/2^+)}$ (right) with respect to the threshold value $\omega_c$, calculated using the charmed baryon multiplet $[\mathbf{6}_F, 2, 0, \rho\lambda]$. The shady bands in these figures are obtained by changing $T$ inside Borel windows.
There exist Borel windows as long as $\omega_c([\Sigma_c, 2, 0, \rho\lambda]) \geq 2.8$ GeV (left), $\omega_c([\Xi_c^\prime, 2, 0, \rho\lambda]) \geq 3.6$ GeV (left) and $\omega_c([\Omega_c, 2, 0, \rho\lambda]) \geq 4.0$ GeV (right).
Accordingly, we choose $\omega_c$ to be around 3.0, 3.7, and 4.0 GeV in the left, middle and right panels, respectively.
}
\label{fig:mass2011}
\end{center}
\end{figure*}

\item We use the baryon multiplet $[\mathbf{6}_F, 3, 1, \rho\rho]$ to perform sum rule analyses, but there do not exist Borel windows when $\omega_c \leq 4.0$ GeV. We also use the baryon multiplet $[\mathbf{6}_F, 1, 1, \lambda\lambda]$ to perform sum rule analyses, but the obtained results depend significantly on the threshold value $\omega_c$. Hence, these results can not be well interpreted, and we shall not use them to draw conclusions.

\end{enumerate}

\section{Summary}
\label{sec:summary}

In this paper we used the method of QCD sum rules within HQET to study the $D$-wave charmed baryons of the $SU(3)$ flavor $\mathbf{6}_F$, and calculated their masses up to the order $\mathcal{O}(1/m_Q)$.
We investigated five charmed baryon doublets, $[\mathbf{6}_F, 1, 1, \rho\rho]$, $[\mathbf{6}_F, 3, 1, \rho\rho]$, $[\mathbf{6}_F, 1, 1, \lambda\lambda]$, $[\mathbf{6}_F, 3, 1, \lambda\lambda]$ and $[\mathbf{6}_F, 2, 0, \rho\lambda]$ (we note that in Ref.~\cite{Chen:2016phw} we failed to construct the currents belonging to the other two multiplets, $[\mathbf{6}_F, 2, 1, \rho\rho]$ and $[\mathbf{6}_F, 2, 1, \lambda\lambda]$).
The results are summarized in Table~\ref{tab:results} for the charmed baryon multiplets $[\mathbf{6}_F, 1, 1, \rho\rho]$, $[\mathbf{6}_F, 3, 1, \lambda\lambda]$, and $[\mathbf{6}_F, 2, 0, \rho\lambda]$, and those obtained using the former two multiplets ($[\mathbf{6}_F, 1, 1, \rho\rho]$ and $[\mathbf{6}_F, 3, 1, \lambda\lambda]$) are reasonable/better. We note that there are large theoretical uncertainties in our mass predictions, but the mass splittings within the same doublet are produced quite well with much less theoretical uncertainty.

Our results suggest that the $\Xi_c(3123)$ observed by the BaBar Collaboration~\cite{Aubert:2007dt} can be well interpreted as a $D$-wave $\Xi_c^\prime$ state, while its quantum numbers can not be determined. It may belong to either the charmed baryon doublet $[\Xi_c^\prime(\mathbf{6}_F), 1, 1, \rho\rho]$ or $[\Xi_c^\prime(\mathbf{6}_F), 3, 1, \lambda\lambda]$, but in both cases there would be a partner state close to it.
Our results also suggest that there may exist as many as four $D$-wave $\Omega_c$ states in the energy region $3.3\sim3.5$ GeV. They are the $D$-wave $\Omega_c$ states of $J^P = 1/2^+$, $3/2^+$, $5/2^+$, and $7/2^+$; the former two belong to the charmed baryon multiplet $[\mathbf{6}_F, 1, 1, \rho\rho]$, and the latter two belong to $[\mathbf{6}_F, 3, 1, \lambda\lambda]$. Recalling that the LHCb experiment observed as many as five excited $\Omega_c$ states~\cite{Aaij:2017nav} at the same time, we propose to search for these $D$-wave $\Omega_c$ states in the further LHCb and BelleII experiments in order to study the fine structure of the strong interaction.
We note that the doubly charmed baryon $\Xi_{cc}^{++}$ was recently discovered by the LHCb Collaboration~\cite{Aaij:2017ueg}, which can also be an idea platform to study the fine structure of the strong interaction~\cite{Chen:2017sbg}.

Following the same procedures, we have investigated the $D$-wave bottom baryons of the $SU(3)$ flavor $\mathbf{6}_F$. The pole mass of the bottom quark $m_b = 4.78 \pm 0.06$ GeV~\cite{pdg} is used, and the obtained results are listed in Table~\ref{tab:bottomresults}. We also suggest to search for them in further experiments. To end this paper, we note that not only masses but also decay and production
properties are useful to understand the heavy baryons~\cite{Chen:2017sci}, and J-PARC is planning an experimental project of such studies~\cite{e50}.
In near future, the joint efforts from experimentalists and theorists will be helpful to identify more and more charmed and bottom baryons.

\begin{table*}[hbt]
\begin{center}
\renewcommand{\arraystretch}{1.6}
\caption{Masses of the $D$-wave bottom baryons of the $SU(3)$ flavor $\mathbf{6}_F$, obtained using the bottom baryon multiplets $[\mathbf{6}_F, 1, 1, \rho\rho]$, $[\mathbf{6}_F, 3, 1, \lambda\lambda]$ and $[\mathbf{6}_F, 2, 0, \rho\lambda]$.
}
\begin{tabular}{c | c | c | c | c c | c c | c | c | c}
\hline\hline
\multirow{2}{*}{Multiplets} & \multirow{2}{*}{B} & $\omega_c$ & Working region & $\overline{\Lambda}$ & $f$ & $K$ & $\Sigma$ & Baryons & Mass & Difference
\\ & & (GeV) & (GeV) & (GeV) & (GeV$^{5}$) & (GeV$^2$) & (GeV$^2$) & ($j^P$) & (GeV) & (MeV)
\\ \hline\hline
\multirow{6}{*}{$[\mathbf{6}_F, 1, 1, \rho\rho]$} & \multirow{2}{*}{$\Sigma_b$} & \multirow{2}{*}{3.2} & \multirow{2}{*}{$0.425< T < 0.487$} & \multirow{2}{*}{$1.425$} & \multirow{2}{*}{$0.079$} & \multirow{2}{*}{$-1.372$} & \multirow{2}{*}{$0.0091$} & $\Sigma_b(1/2^+)$ & $6.28^{+0.18}_{-0.12}$ & \multirow{2}{*}{$2^{+4}_{-2}$}
\\ \cline{9-10}
& & & & & & & & $\Sigma_b(3/2^+)$ & $6.28^{+0.18}_{-0.12}$ &
\\ \cline{2-11}
& \multirow{2}{*}{$\Xi_b^\prime$} & \multirow{2}{*}{3.7} & \multirow{2}{*}{$0.395< T < 0.574$} & \multirow{2}{*}{$1.561$} & \multirow{2}{*}{$0.146$} & \multirow{2}{*}{$-0.961$} & \multirow{2}{*}{$0.0057$}
& $\Xi_b^\prime(1/2^+)$ & $6.39^{+0.11}_{-0.14}$ & \multirow{2}{*}{$1^{+2}_{-1}$}
\\ \cline{9-10}
& & & & & & & & $\Xi_b^\prime(3/2^+)$ & $6.39^{+0.11}_{-0.14}$ &
\\ \cline{2-11}
& \multirow{2}{*}{$\Omega_b$} & \multirow{2}{*}{4.2} & \multirow{2}{*}{$0.421< T < 0.655$} & \multirow{2}{*}{$1.761$} & \multirow{2}{*}{$0.274$} & \multirow{2}{*}{$-1.302$} & \multirow{2}{*}{$0.0042$}
& $\Omega_b(1/2^+)$ & $6.61^{+0.12}_{-0.16}$ & \multirow{2}{*}{$1^{+1}_{-1}$}
\\ \cline{9-10}
& & & & & & & & $\Omega_b(3/2^+)$ & $6.61^{+0.12}_{-0.16}$ &
\\ \hline
\multirow{6}{*}{$[\mathbf{6}_F, 3, 1, \lambda\lambda]$} & \multirow{2}{*}{$\Sigma_b$} & \multirow{2}{*}{3.1} & \multirow{2}{*}{$0.445< T < 0.459$} & \multirow{2}{*}{$1.432$} & \multirow{2}{*}{$0.014$} & \multirow{2}{*}{$-3.139$} & \multirow{2}{*}{--}
& $\Sigma_b(5/2^+)$ & \multirow{2}{*}{$6.38^{+0.43}_{-0.17}$} & \multirow{2}{*}{--}
\\ \cline{9-9}
& & & & & & & & $\Sigma_b(7/2^+)$ & &
\\ \cline{2-11}
& \multirow{2}{*}{$\Xi_b^\prime$} & \multirow{2}{*}{3.3} & \multirow{2}{*}{$0.458< T < 0.485$} & \multirow{2}{*}{$1.509$} & \multirow{2}{*}{$0.018$} & \multirow{2}{*}{$-3.064$} & \multirow{2}{*}{--}
& $\Xi_b^\prime(5/2^+)$ & \multirow{2}{*}{$6.45^{+0.28}_{-0.15}$} & \multirow{2}{*}{--}
\\ \cline{9-9}
& & & & & & & & $\Xi_b^\prime(7/2^+)$ & &
\\ \cline{2-11}
& \multirow{2}{*}{$\Omega_b$} & \multirow{2}{*}{3.5} & \multirow{2}{*}{$0.490< T < 0.509$} & \multirow{2}{*}{$1.612$} & \multirow{2}{*}{$0.024$} & \multirow{2}{*}{$-3.085$} & \multirow{2}{*}{--}
& $\Omega_b(5/2^+)$ & \multirow{2}{*}{$6.55^{+0.22}_{-0.14}$} & \multirow{2}{*}{--}
\\ \cline{9-9}
& & & & & & & & $\Omega_b(7/2^+)$ & &
\\ \hline
\multirow{6}{*}{$[\mathbf{6}_F, 2, 0, \rho\lambda]$} & \multirow{2}{*}{$\Sigma_b$} & \multirow{2}{*}{3.0} & \multirow{2}{*}{$0.398< T < 0.457$} & \multirow{2}{*}{$1.336$} & \multirow{2}{*}{$0.022$} & \multirow{2}{*}{$-2.156$} & \multirow{2}{*}{$0.0082$} & $\Sigma_b(3/2^+)$ & $6.23^{+0.19}_{-0.13}$ & \multirow{2}{*}{$3^{+6}_{-3}$}
\\ \cline{9-10}
& & & & & & & & $\Sigma_b(5/2^+)$ & $6.23^{+0.20}_{-0.13}$ &
\\ \cline{2-11}
& \multirow{2}{*}{$\Xi_b^\prime$} & \multirow{2}{*}{3.7} & \multirow{2}{*}{$0.505< T < 0.543$} & \multirow{2}{*}{$1.793$} & \multirow{2}{*}{$0.076$} & \multirow{2}{*}{$-2.214$} & \multirow{2}{*}{$0.0055$}
& $\Xi_b^\prime(3/2^+)$ & $6.69^{+0.11}_{-0.09}$ & \multirow{2}{*}{$2^{+3}_{-2}$}
\\ \cline{9-10}
& & & & & & & & $\Xi_b^\prime(5/2^+)$ & $6.69^{+0.09}_{-0.09}$ &
\\ \cline{2-11}
& \multirow{2}{*}{$\Omega_b$} & \multirow{2}{*}{4.0} & \multirow{2}{*}{$0.561< T < 0.575$} & \multirow{2}{*}{$1.986$} & \multirow{2}{*}{$0.115$} & \multirow{2}{*}{$-2.121$} & \multirow{2}{*}{$0.0046$}
& $\Omega_b(3/2^+)$ & $6.88^{+0.09}_{-0.08}$ & \multirow{2}{*}{$2^{+2}_{-2}$}
\\ \cline{9-10}
& & & & & & & & $\Omega_b(5/2^+)$ & $6.88^{+0.09}_{-0.08}$ &
\\ \hline \hline
\end{tabular}
\label{tab:bottomresults}
\end{center}
\end{table*}

\section*{ACKNOWLEDGMENTS}

We thank Xian-Hui Zhong and Cheng-Ping Shen for useful discussions.
This project is supported by the National Natural Science Foundation of China under Grants No. 11722540, No. 11475015, No. 11375024, No. 11222547, No. 11175073, No. 11575008, and No. 11621131001;
the 973 program;
the Ministry of Education of China (SRFDP under Grant No. 20120211110002 and the Fundamental Research Funds for the Central Universities);
the Key Natural Science Research Program of Anhui Educational Committee (Grant No. KJ2016A774);
the National Program for Support of Top-notch Youth Professionals.
A.H. is supported by Grants-in-Aid for Scientific Research (Grants No. JP17K05441(C)).

\appendix
\section{Other Sum Rules}
\label{app:sumrule}

In this appendix we list the sum rules for other currents with different quark contents.
\newcounter{mytempeqncnt1}
\begin{figure*}[hbt]
\normalsize
\hrulefill
\begin{eqnarray}
&& \Pi_{\Xi^\prime_c,1,1,\rho\rho} = f_{\Xi^\prime_c,1,1,\rho\rho}^{2} e^{-2 \bar \Lambda_{\Xi^\prime_c,1,1,\rho\rho} / T}
\\ \nonumber &=& \int_{2m_s}^{\omega_c} [ \frac{11}{80640\pi^4} \omega^9 - \frac{51m_s^2}{8960\pi^4} \omega^7 - \frac{m_s\langle \bar q q \rangle}{16 \pi^2} \omega^5 + \frac{3m_s\langle \bar s s \rangle}{40 \pi^2} \omega^5
\\ \nonumber && ~~~~~~ - \frac{19\langle g_s^2 GG \rangle}{3072 \pi^4} \omega^5 + \frac{33m_s^2\langle g_s^2 GG \rangle}{512 \pi^4} \omega^3 - \frac{m_s\langle g_s^2 GG \rangle\langle \bar q q \rangle}{16 \pi^2} \omega - \frac{31m_s\langle g_s^2 GG \rangle\langle \bar s s \rangle}{128 \pi^2} \omega]e^{-\omega/T}d\omega \, ,
\\ && f_{\Xi^\prime_c,1,1,\rho\rho}^{2} K_{\Xi^\prime_c,1,1,\rho\rho} e^{-2 \bar \Lambda_{\Xi^\prime_c,1,1,\rho\rho} / T}
\\  \nonumber &=& \int_{2m_s}^{\omega_c} [ - \frac{59}{2217600\pi^4} \omega^{11} + \frac{29m_s^2}{17920\pi^4} \omega^9 + \frac{m_s\langle \bar q q \rangle}{32\pi^2} \omega^7 - \frac{11m_s\langle \bar s s \rangle}{224\pi^2} \omega^7 + \frac{m_s\langle g_s \bar q \sigma Gq \rangle}{4\pi^2} \omega^5
\\ \nonumber && ~~~~~~ + \frac{69\langle g_s^2 GG \rangle}{35840\pi^4} \omega^7 - \frac{2527m_s^2\langle g_s^2 GG \rangle}{61440\pi^4} \omega^5 - \frac{35m_s\langle g_s^2 GG \rangle\langle \bar q q \rangle}{192\pi^2} \omega^3 + \frac{839m_s\langle g_s^2 GG \rangle\langle \bar s s \rangle}{2304\pi^2} \omega^3
\\ \nonumber && ~~~~~~ - \frac{19m_s\langle g_s^2 GG \rangle\langle g_s \bar q \sigma Gq \rangle}{96\pi^2} \omega]e^{-\omega/T}d\omega\, ,
\\ && f_{\Xi^\prime_c,1,1,\rho\rho}^{2} \Sigma_{\Xi^\prime_c,1,1,\rho\rho} e^{-2 \bar \Lambda_{\Xi^\prime_c,1,1,\rho\rho} / T}
\\ \nonumber &=& \int_{2m_s}^{\omega_c} [ \frac{37\langle g_s^2 GG \rangle}{322560 \pi^4} \omega^7 - \frac{7m_s^2\langle g_s^2 GG \rangle}{3840 \pi^4} \omega^5 + \frac{19m_s\langle g_s^2 GG \rangle\langle \bar s s \rangle}{1152 \pi^2} \omega^3]e^{-\omega/T}d\omega\, .
\end{eqnarray}
\begin{eqnarray}
&& \Pi_{\Omega_c,1,1,\rho\rho} = f_{\Omega_c,1,1,\rho\rho}^{2} e^{-2 \bar \Lambda_{\Omega_c,1,1,\rho\rho} / T}
\\ \nonumber &=& \int_{4m_s}^{\omega_c} [ \frac{11}{80640\pi^4} \omega^9 - \frac{9m_s^2}{1120\pi^4} \omega^7 + \frac{9m_s^4}{80 \pi^4} \omega^5 + \frac{m_s\langle \bar s s \rangle}{40 \pi^2} \omega^5 - \frac{7m_s^3\langle \bar s s \rangle}{8 \pi^2} \omega^3
\\ \nonumber && ~~~~~~ -\frac{19\langle g_s^2 GG \rangle}{3072 \pi^4} \omega^5 + \frac{37m_s^2\langle g_s^2 GG \rangle}{256 \pi^4} \omega^3 - \frac{3m_s^4\langle g_s^2 GG \rangle}{64 \pi^4} \omega - \frac{39m_s\langle g_s^2 GG \rangle\langle \bar s s \rangle}{64 \pi^2} \omega]e^{-\omega/T}d\omega \, ,
\\ && f_{\Omega_c,1,1,\rho\rho}^{2} K_{\Omega_c,1,1,\rho\rho} e^{-2 \bar \Lambda_{\Omega_c,1,1,\rho\rho} / T}
\\ \nonumber &=& \int_{4m_s}^{\omega_c} [ - \frac{59}{2217600\pi^4} \omega^{11} + \frac{7m_s^2}{2880\pi^4} \omega^9 - \frac{7m_s^4}{160\pi^4} \omega^7 - \frac{m_s\langle \bar s s \rangle}{28\pi^2} \omega^7 + \frac{15m_s^3\langle \bar s s \rangle}{16\pi^2} \omega^5 + \frac{m_s\langle g_s \bar s \sigma Gs \rangle}{2\pi^2} \omega^5
\\ \nonumber && ~~~~~~ + \frac{23\langle g_s^2 GG \rangle}{11520\pi^4} \omega^7 - \frac{2153m_s^2\langle g_s^2 GG \rangle}{30720\pi^4} \omega^5 + \frac{23m_s^4\langle g_s^2 GG \rangle}{64\pi^4} \omega^3
\\ \nonumber && ~~~~~~ + \frac{149m_s\langle g_s^2 GG \rangle\langle \bar s s \rangle}{384\pi^2} \omega^3 - \frac{239m_s^3\langle g_s^2 GG \rangle\langle \bar s s \rangle}{192\pi^2} \omega
+ \frac{17m_s\langle g_s^2 GG \rangle\langle g_s \bar s \sigma Gs \rangle}{48\pi^2} \omega]e^{-\omega/T}d\omega\, ,
\\ && f_{\Omega_c,1,1,\rho\rho}^{2} \Sigma_{\Omega_c,1,1,\rho\rho} e^{-2 \bar \Lambda_{\Omega_c,1,1,\rho\rho} / T}
\\ \nonumber &=& \int_{4m_s}^{\omega_c} [ \frac{37\langle g_s^2 GG \rangle}{322560 \pi^4} \omega^7 - \frac{7m_s^2\langle g_s^2 GG \rangle}{1920 \pi^4} \omega^5 + \frac{19m_s\langle g_s^2 GG \rangle\langle \bar s s \rangle}{576 \pi^2} \omega^3]e^{-\omega/T}d\omega\, .
\end{eqnarray}
\hrulefill
\vspace*{4pt}
\end{figure*}

\begin{figure*}[hbt]
\normalsize
\hrulefill
\begin{eqnarray}
\Pi_{\Sigma_c,3,1,\lambda\lambda} = f_{\Sigma_c,3,1,\lambda\lambda}^{2} e^{-2 \bar \Lambda_{\Sigma_c,3,1,\lambda\lambda} / T}
&=& \int_{0}^{\omega_c} [\frac{1}{193536\pi^4}\omega^9 - \frac{13\langle g_s^2 GG \rangle}{46080\pi^4} \omega^5]e^{-\omega/T}d\omega \, ,
\\ f_{\Sigma_c,3,1,\lambda\lambda}^{2} K_{\Sigma_c,3,1,\lambda\lambda} e^{-2 \bar \Lambda_{\Sigma_c,3,1,\lambda\lambda} / T}
&=& \int_{0}^{\omega_c} [-\frac{41}{21288960\pi^4}\omega^{11} + \frac{353\langle g_s^2 GG \rangle}{3870720\pi^4} \omega^7]e^{-\omega/T}d\omega \, .
\end{eqnarray}
\begin{eqnarray}
&& \Pi_{\Xi^\prime_c,3,1,\lambda\lambda} = f_{\Xi^\prime_c,3,1,\lambda\lambda}^{2} e^{-2 \bar \Lambda_{\Xi^\prime_c,3,1,\lambda\lambda} / T}
\\ \nonumber &=& \int_{2m_s}^{\omega_c} [ \frac{1}{193536\pi^4} \omega^9 - \frac{m_s^2}{4480\pi^4} \omega^7 - \frac{m_s\langle \bar q q \rangle}{480 \pi^2} \omega^5 + \frac{m_s\langle \bar s s \rangle}{320 \pi^2} \omega^5
\\ \nonumber && ~~~~~~ - \frac{13\langle g_s^2 GG \rangle}{46080 \pi^4} \omega^5 + \frac{5m_s^2\langle g_s^2 GG \rangle}{3072 \pi^4} \omega^3 - \frac{m_s\langle g_s^2 GG \rangle\langle \bar s s \rangle}{256 \pi^2} \omega]e^{-\omega/T}d\omega \, ,
\\ && f_{\Xi^\prime_c,3,1,\lambda\lambda}^{2} K_{\Xi^\prime_c,3,1,\lambda\lambda} e^{-2 \bar \Lambda_{\Xi^\prime_c,3,1,\lambda\lambda} / T}
\\  \nonumber &=& \int_{2m_s}^{\omega_c} [ - \frac{41}{21288960\pi^4} \omega^{11} + \frac{17m_s^2}{161280\pi^4} \omega^9 + \frac{m_s\langle \bar q q \rangle}{960\pi^2} \omega^7 - \frac{9m_s\langle \bar s s \rangle}{4480\pi^2} \omega^7
\\ \nonumber && ~~~~~~ + \frac{353\langle g_s^2 GG \rangle}{3870720\pi^4} \omega^7 - \frac{151m_s^2\langle g_s^2 GG \rangle}{122880\pi^4} \omega^5 - \frac{5m_s\langle g_s^2 GG \rangle\langle \bar q q \rangle}{1728\pi^2} \omega^3 + \frac{31m_s\langle g_s^2 GG \rangle\langle \bar s s \rangle}{4608\pi^2} \omega^3]e^{-\omega/T}d\omega\, .
\end{eqnarray}
\begin{eqnarray}
&& \Pi_{\Omega_c,3,1,\lambda\lambda} = f_{\Omega_c,3,1,\lambda\lambda}^{2} e^{-2 \bar \Lambda_{\Omega_c,3,1,\lambda\lambda} / T}
\\ \nonumber &=& \int_{4m_s}^{\omega_c} [ \frac{1}{193536\pi^4} \omega^9 - \frac{3m_s^2}{8960\pi^4} \omega^7 + \frac{3m_s^4}{640 \pi^4} \omega^5 + \frac{m_s\langle \bar s s \rangle}{480 \pi^2} \omega^5 - \frac{m_s^3\langle \bar s s \rangle}{24 \pi^2} \omega^3
\\ \nonumber && ~~~~~~ - \frac{13\langle g_s^2 GG \rangle}{46080 \pi^4} \omega^5 + \frac{5m_s^2\langle g_s^2 GG \rangle}{1536 \pi^4} \omega^3 - \frac{m_s\langle g_s^2 GG \rangle\langle \bar s s \rangle}{128 \pi^2} \omega]e^{-\omega/T}d\omega \, ,
\\ && f_{\Omega_c,3,1,\lambda\lambda}^{2} K_{\Omega_c,3,1,\lambda\lambda} e^{-2 \bar \Lambda_{\Omega_c,3,1,\lambda\lambda} / T}
\\ \nonumber &=& \int_{4m_s}^{\omega_c} [ - \frac{41}{21288960\pi^4} \omega^{11} + \frac{3m_s^2}{17920\pi^4} \omega^9 - \frac{27m_s^4}{8960\pi^4} \omega^7 - \frac{13m_s\langle \bar s s \rangle}{6720\pi^2} \omega^7 + \frac{m_s^3\langle \bar s s \rangle}{24\pi^2} \omega^5
\\ \nonumber && ~~~~~~ + \frac{353\langle g_s^2 GG \rangle}{3870720\pi^4} \omega^7 - \frac{131m_s^2\langle g_s^2 GG \rangle}{61440\pi^4} \omega^5 + \frac{7m_s^4\langle g_s^2 GG \rangle}{1152\pi^4} \omega^3
\\ \nonumber && ~~~~~~ + \frac{53m_s\langle g_s^2 GG \rangle\langle \bar s s \rangle}{6912\pi^2} \omega^3 - \frac{m_s^3\langle g_s^2 GG \rangle\langle \bar s s \rangle}{64\pi^2} \omega]e^{-\omega/T}d\omega\, .
\end{eqnarray}
\hrulefill
\vspace*{4pt}
\end{figure*}

\begin{figure*}[hbt]
\normalsize
\hrulefill
\begin{eqnarray}
\Pi_{\Sigma_c,2,0,\rho\lambda} = f_{\Sigma_c,2,0,\rho\lambda}^{2} e^{-2 \bar \Lambda_{\Sigma_c,2,0,\rho\lambda} / T}
&=& \int_{0}^{\omega_c} [\frac{1}{48384\pi^4}\omega^9 - \frac{5\langle g_s^2 GG \rangle}{6912\pi^4} \omega^5]e^{-\omega/T}d\omega \, ,
\\ f_{\Sigma_c,2,0,\rho\lambda}^{2} K_{\Sigma_c,2,0,\rho\lambda} e^{-2 \bar \Lambda_{\Sigma_c,2,0,\rho\lambda} / T}
&=& \int_{0}^{\omega_c} [-\frac{1}{168960\pi^4}\omega^{11} + \frac{227\langle g_s^2 GG \rangle}{1451520\pi^4} \omega^7]e^{-\omega/T}d\omega \, ,
\\ f_{\Sigma_c,2,0,\rho\lambda}^{2} \Sigma_{\Sigma_c,2,0,\rho\lambda} e^{-2 \bar \Lambda_{\Sigma_c,2,0,\rho\lambda} / T}
&=& \int_{0}^{\omega_c} [\frac{\langle g_s^2 GG \rangle}{48384\pi^4} \omega^7]e^{-\omega/T}d\omega \, .
\end{eqnarray}
\begin{eqnarray}
&& \Pi_{\Xi^\prime_c,2,0,\rho\lambda} = f_{\Xi^\prime_c,2,0,\rho\lambda}^{2} e^{-2 \bar \Lambda_{\Xi^\prime_c,2,0,\rho\lambda} / T}
\\ \nonumber &=& \int_{2m_s}^{\omega_c} [ \frac{1}{48384\pi^4} \omega^9 - \frac{m_s^2}{1008\pi^4} \omega^7 - \frac{m_s\langle \bar q q \rangle}{72 \pi^2} \omega^5 + \frac{m_s\langle \bar s s \rangle}{48 \pi^2} \omega^5 - \frac{5m_s\langle g_s \bar q \sigma Gq \rangle}{72 \pi^2} \omega^3
\\ \nonumber && ~~~~~~ - \frac{5\langle g_s^2 GG \rangle}{6912 \pi^4} \omega^5 + \frac{5m_s^2\langle g_s^2 GG \rangle}{432 \pi^4} \omega^3 - \frac{5m_s\langle g_s^2 GG \rangle\langle \bar s s \rangle}{108 \pi^2} \omega]e^{-\omega/T}d\omega \, ,
\\ && f_{\Xi^\prime_c,2,0,\rho\lambda}^{2} K_{\Xi^\prime_c,2,0,\rho\lambda} e^{-2 \bar \Lambda_{\Xi^\prime_c,2,0,\rho\lambda} / T}
\\  \nonumber &=& \int_{2m_s}^{\omega_c} [ - \frac{1}{168960\pi^4} \omega^{11} + \frac{137m_s^2}{362880\pi^4} \omega^9 + \frac{37m_s\langle \bar q q \rangle}{5040\pi^2} \omega^7 - \frac{17m_s\langle \bar s s \rangle}{1344\pi^2} \omega^7 + \frac{m_s\langle g_s \bar q \sigma Gq \rangle}{15\pi^2} \omega^5
\\ \nonumber && ~~~~~~ + \frac{227\langle g_s^2 GG \rangle}{1451520\pi^4} \omega^7 - \frac{3419m_s^2\langle g_s^2 GG \rangle}{552960\pi^4} \omega^5 - \frac{13m_s\langle g_s^2 GG \rangle\langle \bar q q \rangle}{648\pi^2} \omega^3 + \frac{461m_s\langle g_s^2 GG \rangle\langle \bar s s \rangle}{10368\pi^2} \omega^3
\\ \nonumber && ~~~~~~ - \frac{11m_s\langle g_s^2 GG \rangle\langle g_s \bar q \sigma Gq \rangle}{1728\pi^2} \omega]e^{-\omega/T}d\omega\, ,
\\ && f_{\Xi^\prime_c,2,0,\rho\lambda}^{2} \Sigma_{\Xi^\prime_c,2,0,\rho\lambda} e^{-2 \bar \Lambda_{\Xi^\prime_c,2,0,\rho\lambda} / T}
\\ \nonumber &=& \int_{2m_s}^{\omega_c} [ \frac{\langle g_s^2 GG \rangle}{48384 \pi^4} \omega^7 - \frac{m_s^2\langle g_s^2 GG \rangle}{2304 \pi^4} \omega^5 + \frac{5m_s\langle g_s^2 GG \rangle\langle \bar s s \rangle}{864 \pi^2} \omega^3]e^{-\omega/T}d\omega\, .
\end{eqnarray}
\begin{eqnarray}
&& \Pi_{\Omega_c,2,0,\rho\lambda} = f_{\Omega_c,2,0,\rho\lambda}^{2} e^{-2 \bar \Lambda_{\Omega_c,2,0,\rho\lambda} / T}
\\ \nonumber &=& \int_{4m_s}^{\omega_c} [ \frac{1}{48384\pi^4} \omega^9 - \frac{m_s^2}{672\pi^4} \omega^7 + \frac{m_s^4}{48 \pi^4} \omega^5 + \frac{m_s\langle \bar s s \rangle}{72 \pi^2} \omega^5 - \frac{5m_s^3\langle \bar s s \rangle}{18 \pi^2} \omega^3 - \frac{5m_s\langle g_s \bar s \sigma Gs \rangle}{36 \pi^2} \omega^3
\\ \nonumber && ~~~~~~ -\frac{5\langle g_s^2 GG \rangle}{6912 \pi^4} \omega^5 + \frac{5m_s^2\langle g_s^2 GG \rangle}{216 \pi^4} \omega^3-\frac{5m_s\langle g_s^2 GG \rangle\langle \bar s s \rangle}{54 \pi^2} \omega]e^{-\omega/T}d\omega \, ,
\\ && f_{\Omega_c,2,0,\rho\lambda}^{2} K_{\Omega_c,2,0,\rho\lambda} e^{-2 \bar \Lambda_{\Omega_c,2,0,\rho\lambda} / T}
\\ \nonumber &=& \int_{4m_s}^{\omega_c} [ - \frac{1}{168960\pi^4} \omega^{11} + \frac{211m_s^2}{362880\pi^4} \omega^9 - \frac{211m_s^4}{20160\pi^4} \omega^7 - \frac{107m_s\langle \bar s s \rangle}{10080\pi^2} \omega^7 + \frac{181m_s^3\langle \bar s s \rangle}{720\pi^2} \omega^5 + \frac{2m_s\langle g_s \bar s \sigma Gs \rangle}{15\pi^2} \omega^5
\\ \nonumber && ~~~~~~ + \frac{227\langle g_s^2 GG \rangle}{1451520\pi^4} \omega^7 - \frac{2839m_s^2\langle g_s^2 GG \rangle}{276480\pi^4} \omega^5 + \frac{427m_s^4\langle g_s^2 GG \rangle}{13824\pi^4} \omega^3
\\ \nonumber && ~~~~~~ + \frac{253m_s\langle g_s^2 GG \rangle\langle \bar s s \rangle}{5184\pi^2} \omega^3 - \frac{263m_s^3\langle g_s^2 GG \rangle\langle \bar s s \rangle}{3456\pi^2} \omega - \frac{11m_s\langle g_s^2 GG \rangle\langle g_s \bar s \sigma Gs \rangle}{864\pi^2} \omega]e^{-\omega/T}d\omega\, ,
\\ && f_{\Omega_c,2,0,\rho\lambda}^{2} \Sigma_{\Omega_c,2,0,\rho\lambda} e^{-2 \bar \Lambda_{\Omega_c,2,0,\rho\lambda} / T}
\\ \nonumber &=& \int_{4m_s}^{\omega_c} [ \frac{\langle g_s^2 GG \rangle}{48384 \pi^4} \omega^7 - \frac{m_s^2\langle g_s^2 GG \rangle}{1152 \pi^4} \omega^5 + \frac{5m_s\langle g_s^2 GG \rangle\langle \bar s s \rangle}{432 \pi^2} \omega^3]e^{-\omega/T}d\omega\, .
\end{eqnarray}
\hrulefill
\vspace*{4pt}
\end{figure*}

\end{document}